\documentclass[12pt]{article}

\usepackage[paper=letterpaper, margin=1in]{geometry} 
\usepackage{amsmath,amssymb,amsfonts, amsbsy, amsthm}
\usepackage{mathtools}
\usepackage{latexsym}
\usepackage[usenames]{color}
\usepackage{xcolor}

\usepackage{graphics}

\usepackage{hyperref}
\usepackage{natbib}[authoryear]

\usepackage{multicol}
\usepackage{multirow}
\usepackage{enumitem}
\usepackage[utf8]{inputenc}
\usepackage[T1]{fontenc}
\usepackage[scaled]{helvet}

\usepackage[english]{babel}

\usepackage[doublespacing, nodisplayskipstretch]{setspace}
\usepackage{placeins}
\usepackage{enumitem}
\setlist[itemize]{noitemsep, topsep=0pt}
\setlist[enumerate]{noitemsep, topsep=0pt}

\renewcommand{\P}{\operatorname{P}}

\newcommand{\vertiii}[1]{{\left\vert\kern-0.25ex\left\vert\kern-0.25ex\left\vert #1 \right\vert\kern-0.25ex\right\vert\kern-0.25ex\right\vert}}

\newcommand{\etal}{\textit{et al}}
\newcommand{\ie}{\textit{i.e.}}

\newcommand{\uH}{\,\mathrm{H}}

\newcommand{\RR}{\mathbb{R}}

\newcommand{\Ar}{\mathcal{A}}

\newcommand{\bq}{\boldsymbol{q}}

\newcommand{\bW}{\boldsymbol{W}}

\newcommand{\bY}{\boldsymbol{Y}}

\newcommand{\btheta}{\boldsymbol{\theta}}

\newcommand{\beps}{\boldsymbol{\epsilon}}

\newcommand{\bzero}{\boldsymbol{0}}

\theoremstyle{plain}
\newtheorem{thm}{Theorem}
\newtheorem{lem}{Lemma}

\theoremstyle{remark}
\newtheorem{remark}{Remark}
\newtheorem{example}{Example}
\theoremstyle{definition}
\newtheorem{definition}{Definition}
\newtheorem{condition}{Condition}

\usepackage[ruled,vlined]{algorithm2e}

\SetCommentSty{algcomment}

\newcommand{\FDP}{\text{FDP}}
\newcommand{\FDR}{\text{FDR}}

\newcommand{\N}{\mathrm{N}}
\newcommand{\Unif}{\mathrm{Unif}}

\newcommand{\st}{\text{st}}
\newcommand{\wk}{\text{wk}}
\newcommand{\md}{\text{md}}

\newcommand{\bI}{\mathbb I}

\newcommand{\mX}{\mathcal X}

\newcommand{\mT}{\mathcal T}

\newcommand{\mA}{\mathcal A}
\newcommand{\mB}{\mathcal B}
\newcommand{\mC}{\mathcal C}
\newcommand{\mD}{\mathcal D}

\newcommand{\mF}{\mathcal F}
\newcommand{\mG}{\mathcal G}

\newcommand{\mM}{\mathcal M}

\newcommand{\mP}{\mathcal P}
\newcommand{\mQ}{\mathcal Q}
\newcommand{\mR}{\mathcal R}

\newcommand{\mU}{\mathcal U}
\newcommand{\mV}{\mathcal V}
\newcommand{\mW}{\mathcal W}

\newcommand{\csil}{{|S|}}
\newcommand{\sumnull}{{\sum_{S\in\mB_0^{(\ell)}}}}
\newcommand{\sumalt}{{\sum_{S\in\mB_1^{(\ell)}}}}
\newcommand{\sumall}{{\sum_{S\in\mB^{(\ell)}}}}

\newcommand{\dist}{\mathrm{dist}}
\newcommand{\dia}{\mathrm{dia}}

\newcommand{\Rfun}[1]{\texttt{#1}}
\newcommand{\J}{\lceil\gamma_m/ \nu_m \rceil}

\newcommand{\node}{\mathrm{node}}
\newcommand{\feat}{\mathrm{feat}}

\newcommand{\el}{{(\ell)}}
\newcommand{\vD}{\boldsymbol{D}}

\newcommand\Item[1][]{%
  \ifx\relax#1\relax  \item \else \item[#1] \fi
  \abovedisplayskip=0pt\abovedisplayshortskip=0pt~\vspace*{-\baselineskip}}

\makeatletter
\newcommand{\mathleft}{\@fleqntrue\@mathmargin0pt}
\newcommand{\mathcenter}{\@fleqnfalse}
\makeatother

\graphicspath{{./figures/}} 

\title{Distance Assisted Recursive Testing}

\author{Xuechan Li\thanks{Xuechan Li is a PhD candidate  of Biostatistics at Duke University.}
\and Anthony D. Sung\thanks{Anthony D. Sung is an Assistant Professor  of Medicine at Duke University.}
  \and Jichun Xie \thanks{Jichun Xie is an Associate Professor of Biostatistics and Bioinformatics at Duke University.}}

\begin{document}

\maketitle

\begin{abstract}
  In many applications, a large number of features are collected with
  the goal to identify a few important ones among them. Sometimes,
  these features lie in a metric space with a known distance matrix,
  which partially reflects their co-importance pattern. Proper use of
  the distance matrix will boost the power of identifying important
  features. Hence, we develop a new multiple testing framework named
  the Distance Assisted Recursive Testing (DART). DART has two
  stages. In stage 1, we transform the distance matrix into an
  aggregation tree, where each node represents a set of features. In
  stage 2, based on the aggregation tree, we set up dynamic node
  hypotheses and perform multiple testing on the tree. All rejections
  are mapped back to the features. Under mild assumptions, the false
  discovery proportion of DART converges to the desired level in high
  probability converging to one. We illustrate by theory and
  simulations that DART has superior performance under various models
  compared to the existing methods. We applied DART to a clinical
  trial in the allogeneic stem cell transplantation study to identify
  the gut microbiota whose abundance will be impacted by the
  after-transplant care.
\end{abstract}
\newpage

\section{Introduction}

A typical multiple testing problem aims to identify a small number of
important features among many with a controlled false discovery
rate. Sometimes, these features lie in a metric space with known
pairwise distances.  For example, in neuro-imaging studies, the
distance between two neurons can be calculated based on their 3D
location and the brain anatomy structure; in microbiome studies, the
distance between any two amplicon sequence variants (ASVs) can be
calculated based on their evolutionary distance; and in spatial
analysis, the Euclidean distances between two sites can be calculated
via their geometric locations. In these examples, neurons, ASVs, and
geometric locations are features of interest. Very often, important
features tend to cluster with each other. If two features are close in
distance, they are likely to be co-important or co-unimportant.  For
example, in microbiome studies, two evolutionarily close ASVs often
perform similar biological functions. If one is important, the other
is probably important too. Thus when testing the ASV abundance
association with the treatment, if we can properly incorporate their
evolutionary distance, the testing power will be boosted.  In this
paper, we will develop a new multiple testing method which
incorporates the distance information to boost the testing power while
controlling the asymptotic feature-level FDR.

Some existing literature provides alternative solutions to incorporate
distance information into testing.  One of them is to model the
features by hidden Markov chains \citep{sun2009large} or hidden Markov
random fields \citep{Liu2012Graphical-model, shu2015multiple,
  lee2016extended}.  The co-importance patterns are introduced by the
transition probabilities between the importance and unimportance
status among those features. The challenge lies in how to accurately
inferring the transition probabilities. Even assuming all the feature
statistics follow multivariate Gaussian distribution, it is still hard
to derive consistent transition probability estimators without
additional information.
Another solution is to use the weighted or smoothed P-values in the
neighborhood. \citet{zhang2011multiple} developed a method called
FDR$_L$. FDR$_L$ pre-specified a smoothing window. For each
hypothesis, it smooths the p-value across its local neighbors within
the window.  Recently, \citet{cai2020laws} developed a
locally-adaptive weighting and screening method named LAWS. LAWS
weighted the P-value using the estimated local sparsity level, which
is calculated based on a pre-specified kernel function. However, the
performance of LAWS heavily depends on the accuracy in local sparsity
level estimation, while accurately estimating the local sparsity level
is challenging without additional information. In addition, LAWS
focuses on a setting that the features are located in a regular
lattice and require a non-vanishing proportion of important
features. These conditions might not hold for many large-scale feature
selection problems.

In this paper, we propose a new solution called Distance Assisted
Recursive Testing (DART). It embeds multiple testing into an
aggregation tree built upon the feature distances. DART has two
stages.
\begin{itemize}
\item Stage I is to construct an aggregation tree based on the
distance matrix.  First, on layer 1, each node contains only one
feature; it is also called a leaf. On layer $\ell$ ($\ell\geq 2$), we
gradually aggregate the close child nodes from the previous layers to
form new nodes on the current layer. The detailed algorithm is
described in Section~\ref{sec::stgi}.
\item Stage II is to perform multiple testing (of testing feature
  importance) on the aggregation tree from Stage I.  On layer 1, we
  apply the multiple testing procedure to asymptotically control the
  feature-level FDR. Traditional multiple testing method will stop
  after one-layer of testing but DART will not. On layer $\ell$
  ($\ell\geq 2$), the already-rejected child nodes from the previous
  layers will be excluded from the nodes on the current layer to form
  dynamic working nodes. Next, we apply the new multiple testing
  procedure on the working nodes to control the node-level FDR up to
  layer $\ell$. If a node on layer $\ell$ is rejected then all its
  containing features will be rejected. This rejection rule is very
  aggressive but the feature-level FDR will still be asymptotically
  controlled under mild conditions (See
  Section~\ref{sec::theory}). The detailed algorithm is described in
  Section~\ref{sec::stgii}.
\end{itemize}
 The
underlying logic of DART lies in the assumption that closer features
are more likely to have co-importance or co-unimportance
patterns. Some important features could have weak signal-to-noise
ratios. If one such feature stands alone, its chance to be discovered
is hampered by the weak signal-to-noise ratios; if several such
features are aggregated, their collective signal-to-noise ratios will
be amplified, and thus their chances to be discovered are boosted.

Generally speaking, DART is a hierarchical multiple testing
procedure. Some other multiple testing methods also have hierarchical
or graphical structures. \citet{Goeman2012The-inheritance} and
\citet{Meijer2015A-multiple} developed the FWER controlling procedures
on the trees and directed acyclic graphs.
\citet{Dmitrienko2013General} developed methods testing hierarchically
ordered hypotheses with applications to clinical trials and control
FWER.  \citet{Yekutieli2008Hierarchical} considers the case when all
the original hypotheses represent a node on the tree and develop a
method to test those hypotheses simultaneously. Their parent-node
P-values are independent from the child node P-values, very different
from our model. \citet{Guo2018A-New-Approach} developed a per-family
error rate (PFER) and FDR controlling procedure for hypotheses with a
DAG structure.  \citet{Soriano2017Probabilistic} develops a up-down
testing procedure embedded in the partition tree to asymptotically
controls node-level FDR for all nodes on the
tree. \citet{Li2020A-bottom-up} developed a bottom-up multiple testing
approach embedded in the aggregation tree.

Although some existing hierarchical multiple testing procedures share
some similarities with DART, their settings and focuses are very
different. For example, the existing testing methods often assume the
tree structure among nodes are known and static, the node P-values
follow $\mathrm{Unif}(0,1)$ under the null, and aims to control
node-level FDR. DART is very different from the existing testing
methods. The innovations and main contributions of our paper include the
following.
\begin{itemize}
\item First, unlike many existing methods, the tree structure of DART is not
  given but constructed based on the distance matrix via the proposed
  algorithm~\ref{alg:tree}.
\item Second, when testing on this aggregation tree, on higher layers,
  the nodes and hypotheses are dynamic, \ie, depending on the testing
  results on the previous layers. Controlling FDR for dynamic
  hypotheses is challenging. In this paper we introduced new
  techniques to guarantee the asymptotic validity of DART.
\item Third, to make sure DART can be applied to a wide range of
  application contexts, we relaxed the requirement on the input
  feature P-values. P-values obtained from asymptotic tests (such as
  the Wald tests, the score tests, the likelihood ratio tests, \etal)
  often slightly deviate from the uniform or sub-uniform distribution
  though asymptotically they are uniformly or sub-uniformly
  distributed. For multiple testing problems, the slight deviations
  could accumulate and eventually inflate FDR. We proposed the new
  asymptotic oracle P-value definition to guarantee asymptotic FDR
  control while using some of these P-values.
\item Last but not least, we focus on not only the node-level FDR
  control but also the feature-level FDR control. The feature-level
  FDR control is more challenging than the node-level FDR control
  because a node could contain multiple features with mixed
  null/alternative status. We studied the conditions under which the
  feature-level can by asymptotically controlled, which sheds light on
  the appropriate application contexts where DART should be used.
\end{itemize}


The rest of the paper is organized as follows.  Section
\ref{sec::method} describes the DART algorithms.  Section
\ref{sec::theory} justifies the asymptotic validity of DART under mild
conditions.  Section \ref{sec::num} shows that under various models,
DART has superior performance than the competing methods.  Section
\ref{sec::data} applies DART to study the impact hematopoietic stem
cell transplantation (HCT) post-transplant care on patient gut
microbiota compositions. Section \ref{sec::dis} provides a brief
discussion on the possible extension of DART. The proofs of
propositions and theorems are provided in the appendix. More details
on the DART algorithms and the proofs of the lemmas are provided in
the supplementary materials.


\section{Method} \label{sec::method}

\subsection{Model}

Denote by $\Omega = \{1,\ldots,m\}$ the set with $m$ features. Assume
the distance matrix of these features is $\vD = (d_{ij})_{m\times m}$,
where $d_{ij} = d_{ji}$ is the distance between feature $i$ and
feature $j$. It is easy to see that $d_{ii}=0$. The distance matrix
can be scaled so that $\max_{i\neq j} d_{ij} = 1$.

Among these features, let $\Omega_1$ be the important (alternative)
feature set, $\Omega_0$ is the unimportant (null) feature set, and
$\Omega_1 \cap \Omega_0=\emptyset$, $\Omega_1\cup
\Omega_0=\Omega$. For feature $i$, the hypothesis is
\begin{equation}
\label{numhypo}
\uH_{0i}: i \in \Omega_0 \quad \mbox{versus} \quad \uH_{1i}: 
i \in \Omega_1. 
\end{equation}
To test $\uH_{0,i}$, a feature P-value (statistic) $T_i$ is derived.
\begin{definition}[Oracle P-value]
  We call a statistic $\tilde{T}_i$ an oracle P-value if
  \[\P(\tilde{T}_i \leq p ) \leq p \text{ when }  i\in \Omega_0 \quad \text{and}\quad 
  \P(\tilde{T}_i \leq p ) > p \text{ when } i\in \Omega_1.\]
\end{definition}

Under many circumstances, the P-values are derived from the asymptotic
tests (such as the Wald test, the score test, and the likelihood ratio
test), and thus are not oracle P-values; however, they asymptotically
converge to the oracle P-values.

\begin{definition}[Asymptotic oracle P-value]
  We call a statistic $T_i$ an asymptotic oracle P-value if \begin{equation}
  \label{eq:T-conv}
  \sup_{i\in \Omega_0} \sup_{p\in \mP_{i0}} 
\bigg|\frac{\P(T_i < p)}{\P(\tilde T_i < p)}-1\bigg|\leq \delta_{0m} \quad \text{with } \lim_{m\rightarrow
    \infty} \delta_{0m} = 
    o(1),
\end{equation}
where $\mP_{i0} = \left\{ p\in [0,1]: P({\tilde T}_{i} < p) \geq \big\{m(\log m \log \log m)^{1/2}\big\}^{-1} \right\}$.
\end{definition}

In this paper, we assumes all the feature P-values are asymptotic
oracle P-values.  This assumption is easily satisfied by many commonly
used models and tests. Here we provide a linear model example with
features as outcomes. In fact, we used this model to study the impact
of HCT post-transplant care on patient gut microbiota
complications. Please see Section~\ref{sec::data} for details.

\begin{example}
  Consider the linear regression model:
\begin{equation}
  \label{eq::lm}
  \bY_{n\times m} = \bW_{n\times p_0} \btheta_{p_0\times m} +\beps_{n\times m},
\end{equation}
where $\bY_{n\times m} = (\bY_1,\ldots,\bY_m)$ is a feature outcome
matrix with $n$ observations of $m$ features allowing $m>n$,
$\bW_{n\times p_0}$ is the design matrix with $n$ observations of
$p_0$ covariants with $p_0<n$, $\beps_{n\times m}$ is the random error matrix
with $E(\beps)=\bzero$, and
$\btheta_{p_0\times m}= (\btheta_1,\ldots,\btheta_m)$ is the
coefficient matrix with $\btheta_i \in \RR^{p_0}$ the coefficient of
$\bW$ on $\bY_i$.  In many applications, we would like to test
contrasts: for feature $i$, the hypothesis is
$\uH_{0i}:\bq^T\btheta_{i}=0$. We can use the Wald's test to calculate P-values of  $\uH_{0,i}$. Let
$\hat{\btheta}= (\hat{\btheta}_1,\ldots,\hat{\btheta}_m)$ be the
least square estimator of $\btheta$. The Wald's statistic $X_i^\ast$
and its corresponding P-value $T_i$ is
  \begin{equation} \label{eq::lmteststa}
X_i^\ast = \frac{(\bq^T\hat{\btheta}_{i})^2}{s^2 \bq^T
       (\bW^T\bW)^{-1}\bq}, \quad T_i=1-F_0(X_i^\ast),
 \end{equation}
 where
 $s^2 = \frac{1}{n-p_0} \left \lVert \bY_i - \bW\hat{\btheta}_i
 \right\rVert_2^2$, and $F_0$ is the CDF of the $\chi^2(1)$
 distribution.  Here, $T_i$s are not oracle P-values, but they are
 asymptotic oracle P-values. Details are provided in 
 Lemma~\ref{lem:lin-reg-T} and its proof in the supplementary materials.


\end{example}

\subsection{Two stages of DART}

DART has two stages. In stage I, we transform the feature distance
matrix into an aggregation tree where closer features are prioritized
to be aggregated.  In stage II, we embed multiple testing in the
constructed aggregation tree and control the feature-level FDR.
Utilizing trees to incorporate the distance matrix information can
avoid the challenges in estimating the unknown linkage structures
between the distance and the hypothesis status, because the
hierarchical structure of trees automatically leads to the dynamic
exploration of the optimal feature combining levels to adaptively
increase the power.

\begin{figure}[!htbp]
\centering
\includegraphics[scale=0.65]{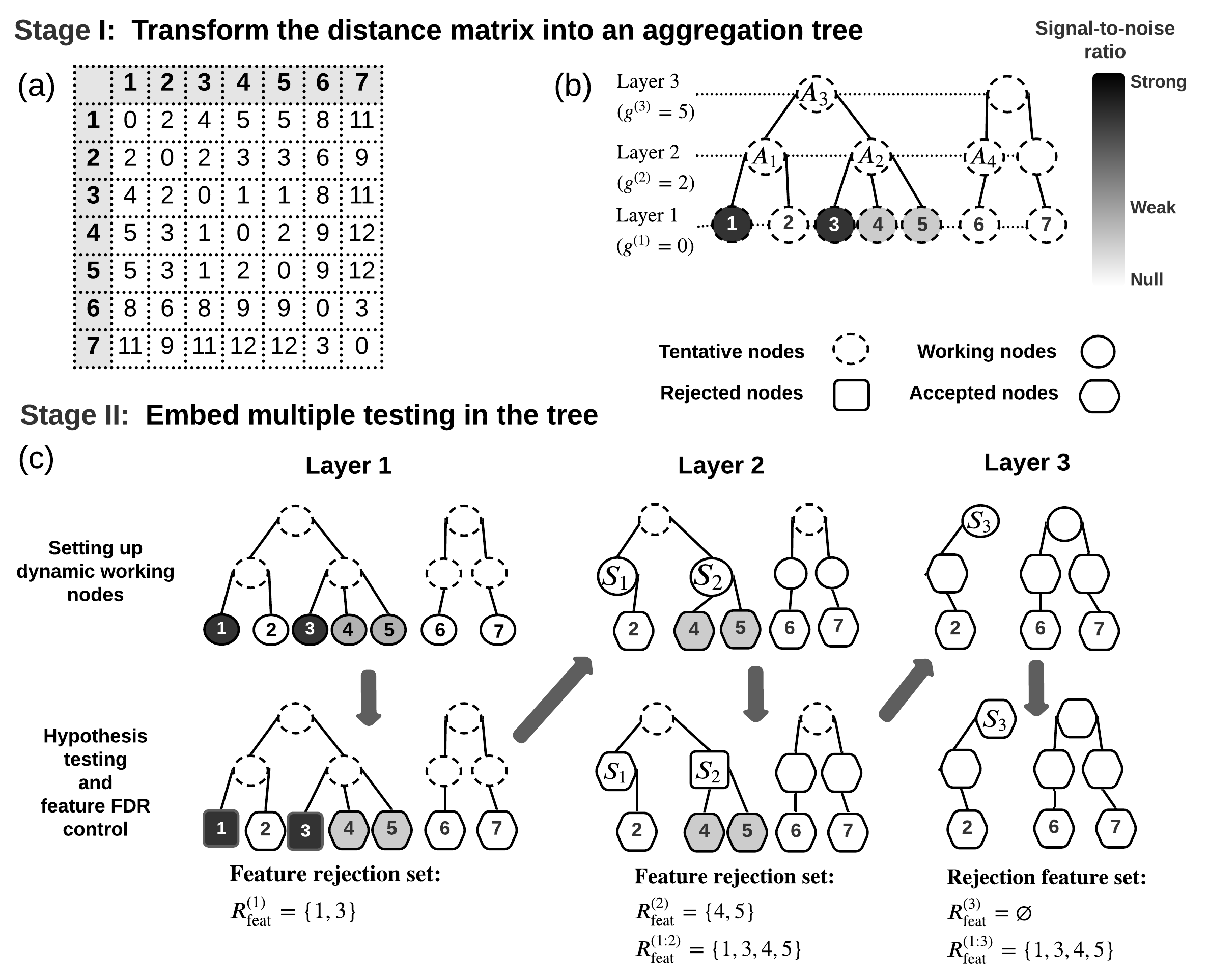}
\caption{An illustrating example of DART with 7 features. (a) Distance
  matrix of the 7 features. (b) In stage I, we transfer the distance
  matrix into the 3-layer aggregation tree based on
  Algorithm~\ref{alg:tree}. The underlying feature signal-to-noise
  ratios are illustrated by the gray scales; these ratios are
  unknown. All nodes at this step are tentative. (c) In stage II, we
  perform the multiple testing embedded in the aggregation tree. We
  start from layer 1 and hierarchically proceed to higher layers. When
  testing on layer $\ell$, all previous rejected features are excluded
  from the temporary nodes (dashed-line circled) to form the working
  nodes (solid-line circled) on this layer. The rejected nodes are
  marked by solid squares and the accepted nodes solid hexagons. All
  the features contained in the rejected nodes are rejected.}
\label{fig:dart-stgs}
\end{figure}

\subsubsection{Stage I: Transform the distance matrix into an
  aggregation tree}\label{sec::stgi}

Before we introduce the tree construction algorithm, we introduce some
notations. Denote an $L$-layer aggregation tree by
$\mT_L=\{\mA^{\el}: \ell=1,\ldots, L\}$, where $\mA^{(\ell)}$ is the
set of nodes on layer $\ell$. Any node $A \in \mA^{(\ell)}$ is a set
of features. If a node $A$ is aggregated from one or multiple children
on layer $\ell-1$, denote its children set by $\mC(A)$. In other
words, $A = \cup_{A'\in \mC(A)} A'$; and $|\mC(A)|$ counts the number
of $A$'s children. For example, in Figure~\ref{fig:dart-stgs}b,
$A_1=\{1,2\}$, $A_2=\{3,4,5\}$, and $\mC(A_3) = \{A_1, A_2\}$ with
$|\mC(A_3)|=2$. A node could be equal to its child. For example, in
Figure~\ref{fig:dart-stgs}, $A_4=\{6\}$ equal to its child. For any
two nodes $A$ and $B$ (not necessarily on the same layer), the
distance between $A$ and $B$ is
$\dist(A,B) = \max_{i\in A, j\in B} d_{ij}$.In
Figure~\ref{fig:dart-stgs}b, $\dist(A_1,A_2) = 5$.  The node distance
defined here can be viewed as the complete linkage function initially
proposed for hierarchical clusterings
\citep{Hastie2009The-elements}. Under some special application
context, other linkage functions may also be used.  For any node $A$,
the diameter of node $A$ is
$\dia(A) = \max_{i\in A, j\in A} d_{ij}$.  In
Figure~\ref{fig:dart-stgs}b, $\dia(A_3) = 5$.

In stage I, we would like to construct an aggregation tree based on
the feature distance matrix. On layer $\ell$ ($\ell\geq 2$), we hope
that for all $A\in \mA^\el$
\begin{equation}\label{eq:node-req}
  \mathrm{dia}(A) \leq g^{(\ell)} \text{ and } |\mC(A)|\leq M.
\end{equation}
The threshold $g^{(\ell)}$ restricts the maximum distance among all
features in the node to make sure its containing features are close to
each other; thus these features are likely to be co-null or
co-alternative. If each of them has weak signal-to-noise ratios,
aggregating them will boost their collective signal-to-noise ratio and
increase their chance to be discovered. We restrict the nodes'
children numbers to reduce the risk of creating mixed nodes (
Definition~\ref{def::mixed}) because too many mixed nodes will
possibly lead to feature-level FDR inflation (see
Section~\ref{sec::theory}). To construct an aggregation tree
satisfying \ref{eq:node-req}, we proposed an algorithm based on the
Greedy algorithm \citep{Cormen2001Introduction}. The pseudo-code of
this stage I algorithm is provided in Algorithm~\ref{alg:tree} in the
the supplementary materials (Section~\ref{sec::alg-stgi}), along with
its remarks.

At the end of stage I, an aggregation tree will be derived, with all
nodes tentative. In stage II, based on the rejection path, we will
further refine those nodes to form working nodes and working
hypotheses.

\subsubsection{Stage II: Embed multiple testing in the tree}
\label{sec::stgii}

The stage II testing procedure is recursive: On layer $\ell$, the
working hypotheses, the working P-values, and the P-value threshold
depend on all previous layers.

On layer 1, leaf $\{i\}$ is coupled with the original hypothesis
$\uH_{0,i}$ in \eqref{numhypo}. We reject $\uH_{0,i}$ if and only if
the working P-value $T_i<\hat{t}^{(1)}(\alpha)$, where $\hat{t}^{(1)}(\alpha)$ is a threshold
defined as follows.
\begin{equation}
  \label{eq:t1}
  \hat{t}^{(1)}(\alpha) = \left\{
    \alpha_{m} \leq t \leq \alpha:
    \frac{mt}{\max\{\sum_{i=1}^m I(T_i<t), 1\}} \leq \alpha,
  \right\}
\end{equation}
where $\alpha_m = 1/\{m(\log m)^{1/2}\}$.  This testing procedure is
similar to the Benjamini and Hochberg procedure
\citep{Benjamini1995Controlling} with minor difference at the
tail. Similar procedure have been proposed and discussed in other
papers such as \citet{liu2013gaussian} and \citet{xie2018false}.
After the testing procedure on layer 1, denote the rejected feature
set by
$R_{\feat}^{(1)} = \{i: T_i\leq \hat{t}^{(1)}(\alpha)\}$.

Traditional multiple testing procedure will stop on layer 1. However,
DART will continue to aggregate nearby nodes because they are likely
to be co-null or co-alternative. If the neighboring nodes all have
weak signal-to-noise ratios, after aggregation their aggregated
signal-to-noise ratio will be larger, and thus their chance to be
discovered will increase..

On layer $\ell$ ($\ell\geq 2$), suppose the testing on
the previous $\ell-1$ layers yields the rejected feature set
$R_{\feat}^{1:(\ell-1)} = \cup_{\ell'=1}^{\ell-1}
R_{\feat}^{(\ell')}$, where $R_{\feat}^{(\ell')}$ is the rejected
feature set on layer $\ell'$.  Denote the tentative node set on layer
$\ell$ of the stage I aggregation tree by $\mA^\el$.  For any
tentative node $A\in \mA^{\el}$, we define
$S(A) = A\setminus R_{\feat}^{1:(\ell-1)}$.  We call $S(A)$ a
\emph{working node}.  The rejected features are removed from the
working nodes because they have already been rejected and do not need
to be tested again. For example, in
Figure~\ref{fig:dart-stgs}c, layer 1 rejected features 1 and 3; on
layer 2, they are removed from the tentative nodes $A_1=\{1,2\}$ and
$A_2=\{3,4,5\}$ to form the working nodes $S_1=\{2\}$ and
$S_2=\{4,5\}$.

Define the testing node set on layer $\ell$:
\[\mB^\el|\mQ^{(1:\ell-1)} = \{S(A): A \in \mA^\el,\ |\mC(S(A))| \geq 2\}.\]
Where $\mC(S(A))=\{A'\setminus R_\feat^{1:(\ell-1)}:A'\in\mC(A)\}$. We exclude the node with only one child because the node must have
been tested on some lower layer. For example, in
Figure~\ref{fig:dart-stgs}c, $S_1=\{2\}$ 
has been tested on layer 1.  For any $S\in \mB^\el$, although it is
dynamic, given the rejection path $\mQ^{(1:\ell-1)}$, they are
deterministic. Thus, conditioning on $\mQ^{(1:\ell-1)}$, we construct
the working node hypotheses:
\[\forall S \in \mB^\el, \quad \uH_{0S}: \ \forall\ j \in S,\ j\in \Omega_0 \quad \text{versus}
  \quad \uH_{1S}: \ \exists\ j \in S,\ j\in \Omega_1,\]

On layer $\ell$, we aim to develop a multiple testing approach to
simultaneously test these (conditional) working node hypotheses while
asymptotically controls the feature-level FDR.

\begin{definition}[Node working P-values]\label{def:agg-P}
  For any node $A$, suppose $T_j$ with $j\in A$ are the feature P-values.
  Then the node's working P-value is defined as 
  \begin{equation}
    X_j = \bar{\Phi}^{-1}(T_j), \quad X_A = \sum_{j\in A} X_j/\sqrt{|A|}, \quad
    T_A = \bar{\Phi}(X_A),
  \end{equation}
where $\bar{\Phi}$ is the complementary CDF of the standard Gaussian
distribution.
\end{definition}

Noteworthy, working P-values are not oracle P-values.
Because $S$ is dynamic, the distribution of $T_S$
depends on $\mQ^{(1:\ell-1)}$. In Lemma~\ref{lem::asym} and \ref{lem::mcond}, we will show $T_S$ still has a good approximation to oracle p-value: \[\sup_{S\in\mB_0^\el}\sup_{p\geq 1/m }\P(T_S\leq p|\mQ^{(1:\ell-1)})\leq p(1+o(1)).\]

Similar to other multiple testing procedures, we threshold the working
P-values to reject the nodes.  For all $S\in \mB^\el$, $\uH_{0S}$ is
rejected if $T_S< \hat{t}^\el (\alpha)$, where
\begin{equation}
  \label{eq:tl}
  \hat{t}^\el(\alpha) = \sup \left\{
    \alpha_m \leq t \leq \alpha: 
    \frac{\sum_{\ell'=1}^{\ell-1} m^{(\ell')} \hat{t}^{(\ell')}(\alpha) +
      m^\el t}{ \max\{ 
      |R_{\feat}^{(1:\ell-1)}| + \sum_{S\in \mB^\el} |S| I(T_S<t), 1
      \}} \leq \alpha
  \right\},
\end{equation}
Here $\alpha_m = 1/\{m(\log m)^{1/2}\}$ and
$m^{(\ell)} = \sum_{S\in \mB^\el} |S|$. For simplicity sake, we use $\hat t^\el$ to present $\hat t^\el(\alpha)$ in the rest of the paper. It is easy to see that
$\hat{t}^\el$ is recursive.

After applying the rejection rule \eqref{eq:tl}, let
\[ \mR_{\node}^{\el} = \{S \in \mB^{\el}:\ T_S < \hat{t}^\el\}, \quad
  R_{\feat}^\el = \cup_{S \in \mR_{\node}^\el}S. \] If a working node
is rejected, We reject all its features. Although this rejection rule
is aggressive, it is reasonable when most close features have
co-null/co-alternative patterns. In Section~\ref{sec::theory}, we will
show this rule asymptotically controls feature-level FDR under mild
conditions. The pseudo-code of the stage II algorithm is provided in
Algorithm~\ref{alg:test} in the supplementary materials
(Section~\ref{sec::alg-stgii}).

\subsection{Tuning parameter selection}\label{sec::tuning}

The number of total layers $L$, the maximum cardinality $M$, and the
distance upper bounds $g^{(2)},...,g^{(L)}$ are viewed as tuning
parameters. Here we provide a feasible approach to select the tuning
parameters.
\begin{itemize}
\item $M=3$. If $M$ is too large, nodes on the aggregation tree are
  more likely to be mixed nodes (Definition~\ref{def::mixed}) and the
  FDR will likely to be inflated. If $M$ is too small, when weak
  signal-to-noise ratio features aggregate, their collective
  signal-to-noise ratios might still be too small to be
  identified. Numerical studies show that $M=3$ performs well in
  practice.
\item $L= \lceil \log_M m - \log_M c_m \rceil$, where $c_m$ is the
  desired minimal number of working nodes on layer $L$. This is
  because on layer $L$, $c_m$ will be lower bounded by
  $m/M^L$.
\item The distance thresholds $g^{(1)},\ldots,g^{(L)}$ are set
  recursively based on the criterion of maximizing the number of
   testable nodes on each layer. Let $g^{(1)}=0$ and
  $G=\{g_1,\ldots,g_K\}$ be the candidate threshold set.  On layer
  $\ell$, let $G^\el=\{g\in G: g>g^{(\ell-1)}\}$. For any
  $g\in G^\el$, let $\mA^\el(g)$ be the resulting node set based on
  Algorithm~\ref{alg:tree}. Then we set $g^\el$ as
  \[g^{(\ell)} = {\arg\max}_{g\in G^\el} |\tilde \mA^\el(g)|, \quad
    \text{where}\ \tilde \mA^\el(g) = \{A: A\in \mA^\el(g),\
    |\mC(A)|\geq 2\}.\] 
\end{itemize}





\section{Asymptotic Theory}
\label{sec::theory}

In this section, we first introduce conditions and theorems to asymptotically control the weighted node-level FDR. Then we discuss how to asymptotically control feature-level FDR. The latter is more challenging.

The common challenges for both parts stem from the dynamic properties
in nodes and node hypotheses, \ie, when testing on layer $\ell$, the
nodes and the testing procedure depend on the testing results on the
previous layers.  Meanwhile, the conditions only describe the
properties of the static features or nodes constructed from stage
I. We developed new techniques to fill in the gap. Specifically, we
carefully analyzed the relationship between the feature signal
strength level and its rejection probability on each layer. By this
way, we can predict the rejection path of some features
probabilistically and based on them to develop the theorems to
asymptotically control the FDRs.

\subsection{Weighted node-level FDR control}

In multiple testing, type I error is commonly measured by the false
discovery proportion (FDP) and its expectation, the false discovery
rate (FDR). Under our model, we defined the weighted node-level FDP
and FDR up to layer $\ell$ as
\begin{equation*}
  \FDP_\node^{(1:\ell)}=\frac{\sum_{\ell'=1}^{\ell}
    \sum_{S\in \mR_{\node}^{(\ell')} \cap \mB_0^{(\ell')} }|S|}
  { \large\{\sum_{\ell'=1}^{\ell} \sum_{S\in \mR_{\node}^{(\ell')}}|S| \large\}\vee 1}. \quad\FDR_\node^{(1:\ell)}=E(\FDP_\node^{(1:\ell)}),
\end{equation*}
Clearly, the denominator of $\FDP_\node^{(1:\ell)}$ counts the
weighted number of all rejected nodes (taking maximum with 1 to avoid
the denominator being 0), and numerator counts the weighted number of
falsely rejected nodes; each node is weighted by its
cardinality. Thus, a larger falsely rejected node will inflate the
weighted node-level FDR more than a smaller falsely rejected node.  We
use the weight node-level FDP and FDR here because it can be more
easily connected with the feature-level FDP and FDR. See
Section~\ref{sec::feature-theory}.

To control the weighted node-level FDR, we introduce the following
conditions.

\begin{condition}\label{cond::sparse}
  Assume $m_1\leq r_2 m^{r_1} \leq r_2 n^{r_1/r_3}$ for some
  $r_1<(M^{L-1}+1)^{-1}$, $r_2>0$, and $r_3>0$.
\end{condition}

Condition~\ref{cond::sparse} assumes the important features are
sparse, and the number of features is bounded by
certain polynomial order of the sample size, $n \geq m^{r_3}$.

For any node $A$, we define its descendant set as
\[\mD(A)=\{D: \exists\ \ell, \text{ such that } D\in \Ar^\el \text{ and } D \subsetneqq A\}.\]
For example, in Figure~\ref{fig:dart-stgs}b,
$\mD(A_3) = \{A_1,A_2, \{1\},\{2\},\{3\},\{4\},\{5\}\}$.

\begin{definition}[Moderately strong Signal-to-Noise Ratio (SNR)
  nodes] A node $A$ is called a moderately strong SNR node if
    \begin{equation}
    \P\{T_A <\alpha_m,\ \forall D\in \mD(A),\ T_D \geq \bar\Phi(m^{r_1-1}\sqrt{\log m}) \} \geq C_1 > 0,\label{eq:mod}
  \end{equation}
  
where $\alpha_m$ is the P-value thresholds lower bound defined in \eqref{eq:tl}.
\end{definition}

In fact, \eqref{eq:mod} is related to the alternative feature SNR.  To better
illustrate the moderately strong SNR nodes, we provide an equivalent
definition when the test statistics follow the Normal distribution.

\begin{example}[Normal distribution example]\label{ex:normal}
  Suppose for feature
$i$, a test statistic $Z_i\sim \mathrm{N}(\tau_i,1)$ can be
derived. The hypotheses are
\[\uH_{0i}: \tau_i=0 \quad \text{versus} \quad \uH_{1i}: \tau_i \neq 0.\]
The P-values are $T_i = 2\bar{\Phi}(|Z_i|)$. 
\end{example}

Under Example \ref{ex:normal}, 
a node A satisfying equation (9) when
\begin{equation*}
  \forall i\in A,\quad |\tau_i|\in [\gamma_m/\sqrt{|A|},\beta_m/\sqrt{|A|-1}].
\end{equation*}
where
\begin{equation}\label{eq:betagam}
  \beta_m=\sqrt{2(1-r_1)\log m-2\log\log m}, \quad \gamma_m=\sqrt{2\log m+\log\log\log m}.
\end{equation}

Although both $\beta_m$ and $\gamma_m$ increase with $m$, the rate is
slow. In practice, when the sample size $n$ increases, $\tau_i$ will
increase with $n$, often at the rate of $\sqrt{n}$. Compared with
$\sqrt{n}$, both $\beta_m$ and $\gamma_m$ are relatively small.

For any moderately strong SNR node $A$, suppose $A\in \Ar^\el$.  We
will prove that with a certain non-vanishing probability, none of
$A$'s descendants will be rejected on the previous layers but $A$ will
be rejected on layer $\ell$. On the tree $\mT_L$, denote the set of
all moderately strong SNR nodes by $\mA_{\text{\md}}$. Define
$c_{\md}=\min_{\ell\in\{1,\ldots,L\}}\lvert \mA_{\md} \cap \mA^\el
\rvert$ as the minimal number of moderately strong SNR nodes across
all layers.

\begin{condition}
  \label{cond::largesignal}
For some constant $r_4>0$, $c_{\md} \geq r_4 \log m$.
\end{condition}

A node on layer $\ell$ has at most $M^{\ell-1}$ features, thus level
$\ell$ has at least $M^{-\ell+1} m_1$ alternative nodes. Because we allow
$m_1=O(m^{r_1})$ by Condition~\ref{cond::sparse}, the total number of
the alternative nodes (containing alternative features) is also
allowed to reach $O(m^{r_1})$. Condition~\ref{cond::largesignal} only
requires $c_\md \geq r_4\log m$ among them are moderately strong SNR
node; therefore, this condition is very weak.

For any node $A$ on the top layer of $\mT_L$, define its dependent node set as
 \begin{equation}
 \label{eq::asyind}
 \Gamma_A=\big\{ A'\in\mA^{(L)}:\{T_i,i\in A\cup A'\}\text{ are dependent}\big\}.
\end{equation}
We assume $\Gamma_A$ is relatively small for most of the $A$s.  We
allow the existence of a) self-dependent nodes whose features are
dependent and b) hub nodes which are dependent with many other nodes,
but these nodes cannot be too many.

\begin{condition}[Few self-dependent and hub nodes]\label{cond::wd}
Define
$\mA'=\{A\in\mA^{(L)}: |\Gamma_A|\geq\delta_{2m}=o(\sqrt{c_\md})\}$. Assume $|\mA'|=o(c_\md)$.
\end{condition}

Under these conditions, the weighted node-level FDP of DART will be
under control and thus also for the weighted node-level FDR.

\begin{thm}[Weighted node-level FDP and FDR control]\label{thm::layerFDR}
  Under Conditions~\ref{cond::sparse}-\ref{cond::wd}, at any
  pre-specified level $\alpha\in (0,1)$, DART satisfies the following
  two statements.
\begin{enumerate}
\item[(1)] For any $\epsilon >0$,
  $\lim_{m,n \to \infty}\P(\FDP_\node^{(1:\ell)} \leq \alpha+\epsilon) =
  1$. Consequently, $\lim_{m,n\to\infty}\FDR_\node^{(1:\ell)}\leq \alpha$.
\item[(2)] Let
  $\tilde{\Omega}_0 = \{j: \tilde{T}_j \text{ follows }
  \mathrm{Unif}(0,1)\}$, where $\tilde{T}_j$ is the oracle P-value of
  feature $j$. If
   \begin{equation}
     \label{eq:null-unif}
     \lim_{m\rightarrow \infty} |\tilde{\Omega}_0|/m = 1, 
   \end{equation}
   then for all $\epsilon>0$,
\[  \lim_{m,n\to\infty} \P(|\FDP_\node^{(1:\ell)}-\alpha| \leq \epsilon) =
  1,\quad \lim_{m,n\to\infty}\FDR_\node^{(1:\ell)}= \alpha.
\]
\end{enumerate}
\end{thm}


\subsection{Feature-level FDR control}\label{sec::feature-theory}

Define the feature-level FDP and FDR up to layer
$\ell$  as
\begin{equation*}
  \FDP_{\feat}^{(1:\ell)}=\frac{| R_{\mathrm {feat}}^{(1:\ell)}\cap \Omega_0|} {|R_{\mathrm {feat}}^{(1:\ell)}|\vee 1}, \quad \FDR_\feat^{(1:\ell)}=E(\FDP_\feat^{(1:\ell)}).
\end{equation*}

It is easy to see that
\begin{align*}
  \sum_{\ell'=1}^{\ell} \sum_{S\in \mR_{\node}^{(\ell')}}|S|
  & =
  \left|\cup_{\ell'=1}^{\ell} \cup_{S\in \mR_{\node}^{(\ell')}} S \right|
  = \left| R_\feat^{(1:l)} \right|\\
  \sum_{\ell'=1}^{\ell} \sum_{S\in \mR_{\node}^{(\ell')} \cap \mB_0^{(\ell')} }|S|
  & =  \left| \cup_{\ell'=1}^{\ell} \cup_{S\in \mR_{\node}^{(\ell')} \cap \mB_0^{(\ell')} } S \right| \leq \left| R_{\mathrm {feat}}^{(1:\ell)}\cap \Omega_0 \right|.
\end{align*}
Thus $\FDP_\node^{(1:\ell)} \leq \FDP_\feat^{(1:\ell)}$. Controlling
$\FDR_\node^{(1:\ell)}$ is easier than controlling
$\FDR_\feat^{(1:\ell)}$. The challenge in controlling
$\FDR_\feat^{(1:\ell)}$ lies in the existence of those nodes
containing both null and alternative features. If such
nodes are rejected, they are counted as true rejections for node-level
weighted FDR control, but the null features in these nodes are counted
as false rejections for feature-level FDR control.

Before we formally define those challenging nodes, we first define the
strong SNR feature set $\Omega_{\st}^{(1:L)}$ and the weak SNR feature
set $\Omega_{\wk}$.

\begin{definition}[Strong SNR feature set]
  Let $\Omega_{\st}^{(1:0)}=\emptyset$.  On layer $\ell$, recursively
  define
\[\mA^{*,\el}=\{A\setminus\Omega_\st^{(1:\ell-1)}:A\in\mA^{\el}\}, \]
and the strong SNR node set as
\begin{equation}\label{eq:stSNR_pv}
  \mG_{\st}^\el  = \left\{S\in \mA^{*,(\ell)}:\ \forall j\in S,\ \P\{T_j \in \kappa(|S|)\} > 1-o(m^{-r_1}) \right\},
\end{equation}
where $\kappa(S)=[m^{-\frac{1-r_1}{|S|-1}},\big\{m(\log m\log\log m)^{1/2}\big\}^{-1/|S|}]$.
Then the strong SNR feature set on layer $\ell$ and up to layer $\ell$ are 
\[
  \Omega_{\st}^{(\ell)} = \cup_{S\in\mG_{\st}^\el}S, \quad
  \Omega_{\st}^{(1:\ell)}
  =\cup_{\ell'=1}^\ell\Omega_{\st}^{(\ell')}.
\]
\end{definition}

Under Example \ref{ex:normal}, $\P\{T_j \in \kappa(|S|)\} > 1-o(m^{-r_1})$ in \eqref{eq:stSNR_pv} is
satisfied when
\[|\tau_j|\in\bigg(\frac{\gamma_m}{\sqrt{|S|}}+\lambda_m,\frac{\beta_m}{\sqrt{|S|-1}}-\lambda_m\bigg),\]
where $\lambda_m=\sqrt{2r_1\log m}$.

We will prove that with a high probability converging to 1, none of
the features in $\cup_{S\in\mG_\st^{(\ell)}}S$ will be rejected from
layer $1$ to layer $\ell-1$ but all of them will be rejected on layer
$\ell$.


\begin{definition}[Weak SNR feature set]
  Let $\iota = (0, m^{\frac{r_1-1}{M^{L-1}}})$. Define the weak SNR feature set
  \begin{equation}\label{eq:wkSNR_pv}
    \Omega_{\wk} = \left\{j\in \Omega_1: \P(T_j\in \iota) =
    o(m^{-r_1}) \right\}.
  \end{equation}
\end{definition}

Under Example \ref{ex:normal}, $\P(T_j\in \iota) =
    o(m^{-r_1})$ in \eqref{eq:wkSNR_pv}
is satisfied when \[|\tau_j|\in(0,\beta_m/\sqrt{M^{L-1}}).\]

When a node $S$ contains only null features and weak signal features,
then the probability of rejecting $S$ is negligible.

\begin{definition}[mixed nodes]
\label{def::mixed}
For any node $A \in \mA^\el$, let
\begin{equation}
  \label{eq:S1-S2}
  A^\ast = A \setminus (\Omega^{(1:\ell-1)}_{\st}\cup\Omega_{\wk}),
\quad A_0^\ast = A^\ast \cap \Omega_0, \quad  \quad A_1^\ast = A^\ast \cap \Omega_1 .
\end{equation} 
If  $A_0^\ast \neq \emptyset$ and $A_1^\ast\neq \emptyset$,  we call $A$ a mixed node.
\end{definition}

Noteworthy, not all nodes containing both null and alternative
features are called mixed nodes. For example, suppose node
$A\in \Ar^{(\ell)}$ have three child nodes $\mC(A)=\{A_1,A_2,A_3\}$,
where $A_1 \subset \Omega_{\st}^{(\ell-1)} $ ,
$A_2 \subset \Omega_{\wk}$, and $A_3$ contains all null
features. Although both null and alternative features exist in node
$A$, this is not a mixed node. This is because $A_1$ will be rejected
on layer $\ell-1$ with high probability converging to 1 so that $A$'s
corresponding working node $S$ is probably $A_2\cup A_3$; also, $S$
will not be rejected on layer $\ell$ with probability converging to 1
so that FDR will not be inflated. Define the strong and weak feature
set will further narrow down the mixed nodes so that the condition to
restrict their number (Condition~\ref{cond::lip}) becomes weaker.

\begin{condition} [Sparse mixed nodes]
\label{cond::lip} 
Let $\mP=\{S\in \mA^{(L)}: S \text{ is a mixed node}\}$. Then $|\mP| = o(c_\md)$.
\end{condition}

Condition \ref{cond::lip} assumes that the mixed nodes on layer $L$
(the top layer) are rare. Equivalently, this means the dominating
majority of the nodes contain: 1) only null features; 2) only
alternative features; 3) a combination of null and alternative
features but all alternative features are either weak or strong SNR
features. Because the aggregation tree is constructed based on the
distance matrix, this condition can be translated as how distance
informs hypothesis states (null or alternative). To prove the
consistence of the overall $\FDP$, we need this condition because our
rejection rule aggressively rejects all features in a node if the node
is rejected. Without Condition \ref{cond::lip} we might reject too
many mixed nodes so that the feature-level FDR could be inflated.

\begin{thm} [Overall feature FDR control]\label{thm::overallFDR}
  Under Conditions~\ref{cond::sparse}-\ref{cond::lip}, at any
  pre-specified level $\alpha\in (0,1)$, DART satisfies the following
  two statements.
\begin{enumerate}
\item[(1)] For any $\epsilon>0$,
  $\lim_{m,n\to \infty}\P(\FDP_\feat^{(1:\ell)} \leq \alpha + \epsilon) =
  1$. Consequently, $\lim_{m,n\to\infty}\FDR_\feat^{(1:\ell)} \leq \alpha$.
\item[(2)] If \eqref{eq:null-unif} holds, then for any $\epsilon>0$, 
\begin{equation*}
\lim_{m,n\to \infty}\P(|\FDP_\feat^{(1:\ell)}-\alpha|\leq\epsilon) =  1, \quad \lim_{m,n\to\infty}\FDR_\feat^{(1:\ell)}=\alpha.
\end{equation*}
\end{enumerate}
\end{thm}


\section{Numerical results}
\label{sec::num}

In this section, the simulation results are carried out to evaluate
the performance of DART.  We simulate $m$ features located in the
two-dimensional Euclidean space with randomly generated location
coordinates: the first coordinate follows $\N(0,2)$, and the second
coordinate follows $\Unif(0,4)$. A distance matrix
$\pmb D=(d_{i,j})_{m\times m}$ is calculated based on the feature
location coordinates.  Two different feature settings are considered,
$m=100$ and $m=1000$. When $m=100$, we generate $m_1=22$ alternative,
and $n=90$ samples. When $m=1000$, we generate $m_1=141$ alternatives,
and $n=300$ samples.

Based on the tuning parameter selection criterion in Section 2.3, we
construct a 2-layer aggregation tree when $m=100$ and a 4-layers
aggregation tree when $m=1000$. More details about the tuning
parameters settings and their selection procedure are shown in
supplementary materials \ref{sec::tuning}.

We consider five different model settings, SE1--SE5.

Throughout the five settings, the hypotheses are:
\[H_{0,i}: \theta_i=0 \quad\text{against}\quad H_{1,i}:\theta_i\neq
  0,\quad i\in\Omega. \]
SE1 simulates the working P-values satisfying
the oracle P-value property, and thus mimics the ideal situation. SE2
and SE3 simulates the working P-values by mis-specifying the null
distributions, and thus these P-values do not satisfy the oracle
P-value property. We use these two settings to evaluate the robustness
of DART and the competing methods. SE4 simulates the linear regression
model and SE5 simulates the Cox proportional hazard model. The feature
P-values are derived from the Wald tests. We are interested to see how
DART compares to the competing methods under these two commonly used
models.  Details in how to generate these simulation settings are
displayed in the supplementary materials
(Section~\ref{sec::simu_settings}). Under each setting, the simulation
is repeated $200$ times. The R codes are available at
\url{https://github.com/xxli8080/DART_Code}.

\begin{figure}[!htbp]
\centering
\includegraphics[scale=0.085]{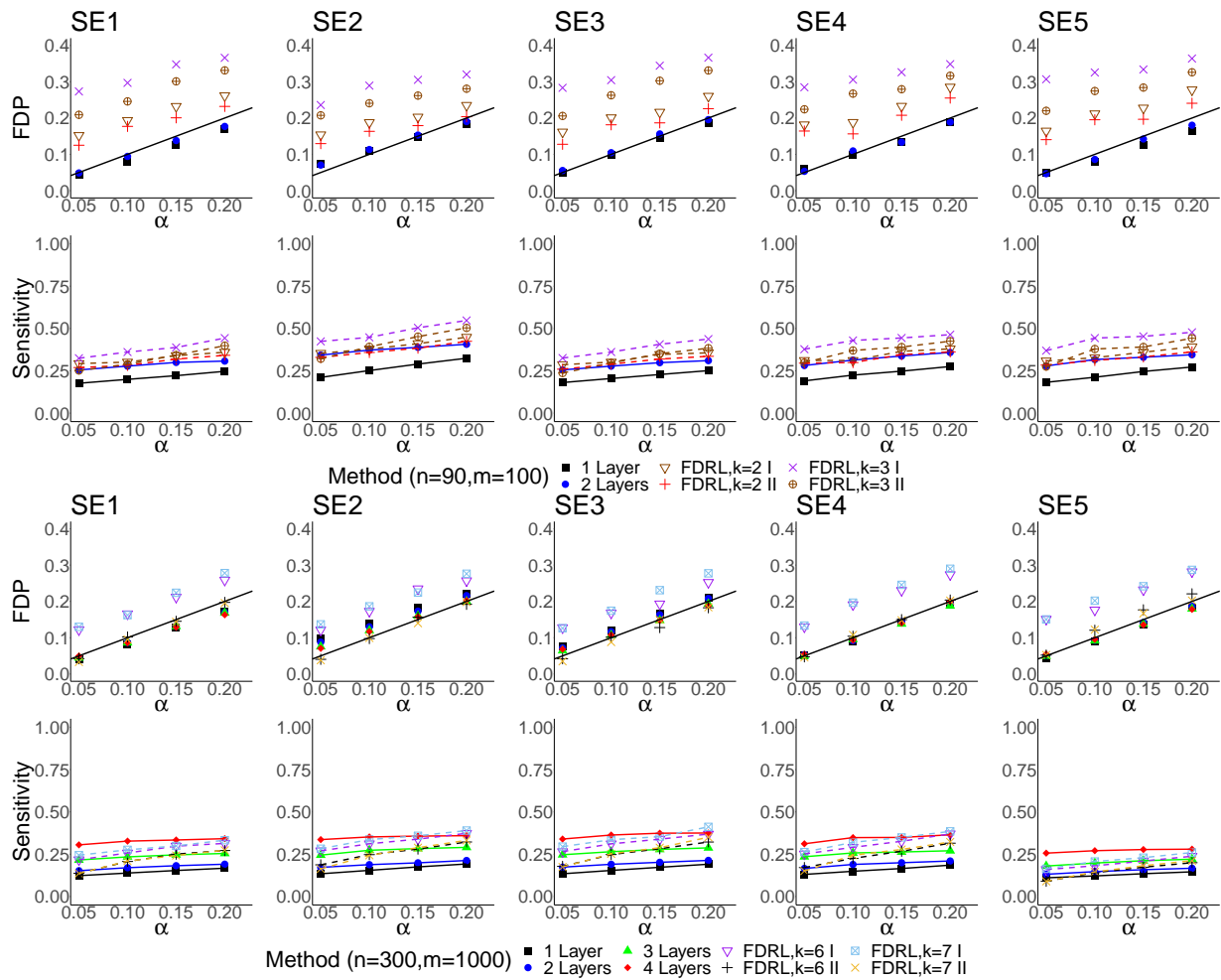}
\caption{Simulation results for setting SE1-SE5. The first two
  rows represent the results in the setting $(n,m)=(90,100)$, and the
  second two rows represent the results in the setting
  $(n,m)=(300,1000)$.}
\label{fig::simple}
\end{figure}

We set the nominal FDR at the level $\alpha=0.05,0.1,0.15,0.20$. We
followed Section~\ref{sec::tuning} to select the tuning parameters;
details are displayed in the supplementary materials
(Section~\ref{sec::simu_tuning}).  DART successfully controlled the
empirical FDR under the desired level. The FDR control is robust when
the model is misspecified.  Figure~\ref{fig::simple} shows how DART
performs when the algorithm stopped at different layers. Obviously,
the one-layer DART is the same as the traditional single layer
multiple testing method which ignores the distance matrix. As the
number of maximum layers $L$ goes up, more alternative features are
aggregated and identified. Notably, increasing the nominal FDR level
cannot lead to such great increase in sensitivity.

We also compared the performance of DART with the two FDR$_L$
procedures (FDR$_L$ I and FDR$_L$ II) proposed by
\citet{zhang2011multiple}. The two procedure adjust each feature's
P-value according to its $k$-nearest neighbors; the adjusted p-value
is the median of its neighborhood p-values. The FDR$_L$ I and FDR$_L$
II procedures use different methods to estimate the distribution of
the adjusted P-values. We compared to both procedures in our
simulation.  To perform a fair comparison, we also tried a wide range
of choices $k\in\{2,3,\ldots,9\}$. When $(n,m)=(90,100)$, both FDR$_L$
I and FDR$_L$ II led to FDR inflation regardless of the choice of
$k$. When $(n,m)=(300,1000)$, the FDR$_L$ I procedure constantly led
to inflated FDR, while the FDR$_L$ II procedure led to the desired FDR
when $k\leq 7$ but the sensitivity is lower than DART.
Figure~\ref{fig::simple} presents the performance of FDR$_L$
procedures with $k=2,3$ when $(n,m)=(90,100)$ and $k=6,7$ when
$(n,m)=(300,1000)$, because under these $k$ settings, the FDR$_L$
procedures perform the best.

One reason that the $\FDR_L$ procedures did not perform as well
as DART is that the methods use a constant $k$ to aggregate P-values.
Very often, the distance among features often cannot be fully captured
by the neighborhood with constant number of neighbors. For example, an
feature far away from all other features also have $k$ nearest
neighbors; however, the isolated feature and its neighbors often do
not share co-importance. Thus, $\FDR_L$ does not perform
well under these settings.

\section{Data Analysis}
\label{sec::data}

We apply DART to a clinical trial on the hematopoietic stem cell
transplantation (HCT), where microbiome data are collected from $144$
leukemia patients before and after the HCT.
Graft-versus-host disease
(GVHD) is one of the major complications of the HCT. Recent studies
have linked GVHD to the disruptions of the gut microbiome
\citep{jenq2012regulation}, and the disruptions may be related to the
environmental changes such as post-transplant care
\citep{claesson2012gut}.  The goal of this study is to investigate the
potential impact of the post-transplant care (home care versus
standard hospital care) on the patient gut microbiota composition.

To achieve the goal, the patient fecal samples are collected before
and after HCT; the fecal microbiome are sequenced by the 16S ribosomal
RNA sequencing at the Memorial Sloan Kettering Cancer Center. The data
are then pre-processed by the R package, \Rfun{DADA2}
\citep{callahan2016dada2}, to generate the amplicon sequence variants
(ASV) and the read counts. Samples with less than $200$ total read
counts and the ASVs with read counts fewer than $4$ in more than
$80\%$ of the samples are removed from the analysis.  After the
pre-processing procedure, the data set contains $288$ samples (before-
and after- HCT) from $144$ patients, each with $97$ ASVs.  The data
are available at
\url{https://github.com/xxli8080/DART_Code/tree/master/Data_Analysis}.
In our analysis, to increase computation stability, the zero counts
are replaced by $0.5$ \citep{aitchison,kurtz2015sparse}.

In microbiome studies, the ASV abundance compositions are more
meaningful than the absolute read counts.  To modeling the
compositional microbiome data, we use the additive log-ratio
transformation proposed by \citet{aitchison}. Specifically, we choose
the most abundant ASV (the ASV with the largest median read counts
across all patients) as the reference ASV, and define $M_{i}$ as the
log read counts ratio between the ASV $i$ and the reference ASV. For
example, for a patient, if the read counts of ASV $i$ and the
reference ASV are $100$ and $200$ respectively, then
$M_i = \log(100/200) = -\log 2$.

Because one ASV is chosen as the reference ASV, the distance matrix is
calculated among the remaining $96$ non-reference ASV using the R
package \Rfun{Phangorn} \citep{phangorn} based on the JC69 model
\citep{jukes1969evolution}.  The JC69 model is a classical Markov
model of DNA sequence evolution and can be used to estimate the
evolutionary distance between sequences.  Two ASVs with similar
sequences tend to be close with each other, and more likely to perform
similar biological functions. Therefore, we will incorporate the
distance matrix in identifying the important ASVs. We use the linear
model, defined in \eqref{eq::lm} with $p_0=3$, to regress the
microbiome composition changes before and after HCT on the
after-transplantation care (home care vs. hospital care) and other
covariates. Specifically, for the non-reference ASV $i$,
$i\in\{1,\ldots,96\}$,
\begin{equation}
\label{eq::linear}
M_{1,i}-M_{0,i} = \theta_{1,i}W_1+\theta_{2,i} W_2 +\theta_{3,i} W_3+\epsilon_i
\end{equation}
Here, $M_{0,i}$ (and $M_{1,i}$) is the log counts ratio between ASV
$i$ and the reference ASV before (and after) the transplant. Thus
$M_{1,i}-M_{0,i}$ is the corresponding $Y_i$ in the model
\eqref{eq::lm}. In addition, $W_1=1$ is the intercept term, $W_2$ is
the type of care, $W_3$ is the length of the care (the gap between the
HCT surgery and the after-care sample collection), and the
$\epsilon\sim N(0,\sigma^2)$ is the random error term with unknown
$\sigma^2$.  To check whether the after-transplant care affects the
ASV compositions, we set up the hypotheses $H_{0i}:\theta_{2,i}=0$,
$i=1,\ldots,96$.  The P-value $T_i$ are calculated based on the Wald
tests.

Based on the tuning parameter selection procedure described in Section
2.3, we construct an aggregation tree with $M=3$,
$L=\lceil\log_M 96-\log_M 30\rceil=2$, and
$g^{(2)}=8/\sqrt{144\log 96\log\log 96}$. The aggregation tree has
$33$ non-single-child nodes on the second layer. The nominal FDR level
is set at $0.1$.

\begin{figure}[!htbp]
\centering
\includegraphics[trim={0 0cm 0 0},clip,scale=0.8]{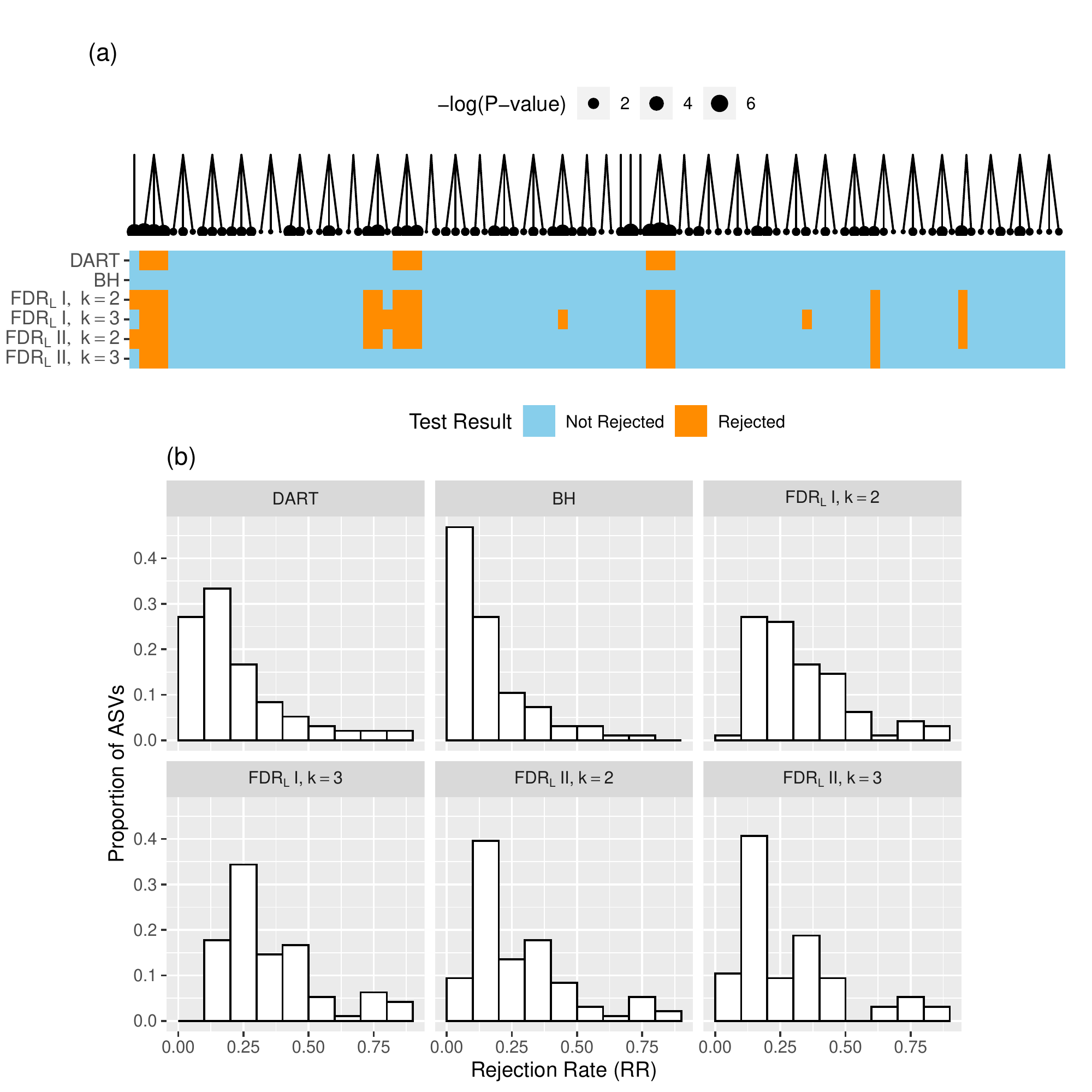}
\caption{(a) Illustration of the leaf P-values, the aggregation tree,
  and the testing results. The leaf size is scaled according to
    the inverse of the P-values. The testing results of different
  methods are shown in different rows, with blue blocks representing
  the accepted non-reference ASVs and orange blocks representing
  the rejected non-reference ASVs.  (b) Histograms of ASV rejection rate
  across the bootstrap with $200$ re-samplings.}
\label{fig::detect}
\end{figure}

The performance of the DART is compared with two competing methods: 1)
BH procedure; 2) FDR$_L$. For the FDR$_L$ I and II procedures, we
considered $k=2$ or $3$. Figure \ref{fig::detect}(a) shows that the
ASVs that are close to each other tend to have similar (small or
large) P-values. This suggests that the co-importance pattern among
similar ASVs might hold here. In the end, the two-layer DART
identified $9$ important ASVs while the traditional BH procedure did
not identify any ASV. Both FDR$_L$ I and FDR$_L$ II procedures
identified $14$ important ASVs when $k=2$. When $k=3$, FDR$_L$ I
identified $16$ important ASVs, and FDR$_L$ II identified $7$
important ASVs.

In order to evaluate the stability of these methods, we conduct the
bootstrap with $200$ re-samplings.  For a specific testing method, the
rejection rate of an ASV is calculated as the ratio of the times that
the ASV is identified in the $200$ rounds of resamplings. If a method
is stable, an ASV should tend to be consistently rejected or
accepted. In other words, for a valid and powerful test, most null
ASVs are expected to have small rejection rates, and very few
alternative ASVs are expected to have high rejection rates.
Figure \ref{fig::detect}(b) shows that DART and BH procedures
generates the histograms with a peak rejection rate within $[0,0.2)$,
while FDR$_L$ have the peak rejection rate between $[0.1,0.3)$. Table
~\ref{tab::toyeg} listed the proportion of ASVs with large ($> 0.8$)
or small ($\leq 0.1$) rejection rates for each method.  Compared with
FDR$_L$ method, DART and BH have a higher proportion of ASVs with
small rejection rates, indicating both DART and BH have lower risk in
FDR inflation. Meanwhile, FDR$_L$ methods have a small proportions of
ASVs with the rejection rate within $0-0.1$, indicating it is not
stable in accepting null ASVs.  On the other hand, DART also has a
higher proportion of ASVs with large rejection rates comparing to the
BH method. This indicates that DART has a robust high power.

\begin{table}[!htbp]
\centering
\caption{Summary of the boostrap results. (RR stands for rejection rates)}
\label{tab::toyeg}
\begin{tabular}{lll}
  \hline
Method & RR$\leq 0.1$ & RR$>0.8$ \\ 
  \hline
DART & $0.27$ & $0.02$ \\ 
BH & $0.47$ & $0$ \\ 
FDR$_L$ I, $k=2$ & $0.01$ & $0.03$ \\
FDR$_L$ I, $k=3$ & $0$ & $0.04$ \\
FDR$_L$ II, $k=2$ & $0.09$ & $0.02$ \\
FDR$_L$ II, $k=3$ & $0.1$ & $0.03$ \\ 
\hline
\end{tabular}
\end{table}


\section{Discussion}
\label{sec::dis}

In this paper, we developed a novel multiple testing method, DART, to
incorporate feature distance in multiple testing. Under many
application contexts, the feature distances serve as auxiliary
information of their co-importance pattern. DART utilizes this
information to boost the testing power. DART applies to the P-values
obtained from many asymptotic tests, and thus can work with a wide
range of models.


Stage 1 of DART involves constructing an aggregation tree. We provided
Algorithm \ref{alg:tree} to construct the aggregation tree.  Other
algorithms may also work, and result in a different aggregation tree
from the same distance matrix. Consequently, Stage 2 testing process
could lead to different results based on different trees. In practice,
if several aggregation trees exist, DART can be applied to all of
them, and we can take the one with the most rejections. The asymptotic
validity will still hold for this procedure.


DART is a multiple testing method embedded in a hierarchical tree that
constructed from the distance matrix. It can be easily extended to the
case where other information implies the co-importance pattern of the
features. Such information could from domain knowledge, external data
sets, or other resources. In addition, the hierarchical testing ideas
and techniques can also be extended to solve other multiple testing
problems.

\appendix

\section*{Appendix: Proof of the Main Theorems}

\label{sec::main_proof}
\newcommand{\tjl}{\theta_{j_i^{(\ell)}}}
Before the proof, we need to introduce some further notations. On layer $\ell$,
for a working node $S\in\mB^{(\ell)}$, let
$\mU(S)=\{S'\subset S:S'\in \cup_{\ell'=1}^{\ell-1}\mB^{(\ell')}\}$
be the collection of sets in the testing path of $S$.
In addition, let $\mU^c(S)=\{S''\in \cup_{\ell'=1}^{\ell-1}\mB^{(\ell')} :S''\cap S=\emptyset, S''\cup S\subset A,\text{ for some }A\in\mA^\el\}$ be the collection of sets that was planning to combined with $S$ on layer $\ell$ of the static aggregation tree but rejected on previous layers. When $S\in\mB^{(1)}$, we set $\mU(S)=\mU^c(S)=\emptyset$. 
We define $G_S(c)$ as the complementary CDF conditional on previous testing results. When $\ell=1$, we have $S=\{i\}\subset\{1,...,m\}$, and $G_S(c)=P(Z_i\geq c)$ with $Z_1,\ldots,Z_m\overset{iid}{\sim} N(0,1)$. When $\ell>1$, the oracle rejection path for set $S\in\mB^\el$ is recursively defined as \[\mQ_z^{(1:\ell-1)}=\{z:\forall S'\in\mU(S), G_{S'}(Z_{S'})\geq\hat t^{(\ell_{S'})}(\alpha),\forall S''\in\mU^c(S), G_{S''}(Z_{S''})\leq\hat t^{(\ell_{S''})}(\alpha)\},\] where
\[
G_S(c)=\P\big(Z_S\geq c\big|\mQ_z^{(1:\ell-1)}\big)
\]
and $Z_S=\sum_{i\in S} Z_i/\sqrt{|S|}$, and $\ell_{S'},\ell_{S''}\in\{1,...,\ell-1\}$ is the value s.t. $S'\in\mB^{(\ell_{S'})}$ and $S''\in\mB^{(\ell_{S''})}$, respectively. \par
Given $Z_1,\ldots,Z_m$ are mutually independent, we have
\[G_S(c)=\P\big(Z_S\geq c\big|\forall S'\in\mU(S), G_{S'}(Z_{S'})\geq\hat t^{(\ell_{S'})}(\alpha)\big)
\]
Given the definition of $G_S(c)$, we define the rejection path as
\begin{equation}
\label{eq::rejp}
\mQ^{(1:\ell-1)}=\{x:\forall S'\in\mU(S), G_{S'}(X_{S'})\geq\hat t^{(\ell_{S'})}(\alpha),\forall S''\in\mU^c(S), G_{S''}(X_{S''})\leq\hat t^{(\ell_{S''})}(\alpha)\}
\end{equation}
\par
In addition, for two sequence of real numbers $a_m$ and $b_m$, we write $a_m=o(b_m)$ when $a_m/b_m\to 0$, and $a_m=O(b_m)$ when $\lim_{m\to\infty}|a_m/b_m|\leq C$ for some constant $C$.
To prove the asymptotic properties of DART, we need the following lemmas.

\begin{lem}\label{lem:lin-reg-T}
  Under the linear regression model \eqref{eq::lm},  $T_i$s are asymptotic oracle P-values.
\end{lem}

\begin{lem}
\label{lem::asym}
Let $\mP_i=\{p\in[0,1]:\P(\tilde T_i<p)\geq \epsilon(m)\}$ and $\mP'_i=\{p\in[0,1]:\P(\tilde T_i<p)\geq \epsilon(m)\epsilon'(m)\}$, with $\epsilon(m), \epsilon'(m)\to 0$. For any set of independent random variable $\hat T_i\in [0,1]$, and a collection $\mM= \{S\subset \{1,...,m\}:|S|<c_0\}$ with some constant $c_0$,
\begin{enumerate}[label=(\arabic*)]
\item If $\quad\max_{i\in \mM}\sup_{p\in \mP'_i}\big|\P(\hat T_i<p)/\P(\tilde T_i<p)-1\big|\to 0$,
then,
\begin{align*}
\sup_{S_0\in\mM}\sup_{p\geq\epsilon(m)}\bigg|\frac{\P(\sum_{i\in S_0}\hat X_i>c_{S_0}(p))}{\P(\sum_{i\in S_0}\tilde X_i>c_{S_0}(p))}-1\bigg|\to 0,
\end{align*}
\item If $\quad\lim_{m\to\infty}\max_{i\in \mM}\sup_{p\in \mP'_i}\big(P(\hat T_i<p)/\P(\tilde T_i<p)-1\big)\leq 0$,
then,
\begin{align*}
\lim_{m\to\infty}\sup_{S_0\in\mM}\sup_{p\geq\epsilon(m)}\bigg(\frac{\P(\sum_{i\in S_0}\hat X_i>c_{S_0}(p))}{\P(\sum_{i\in S_0}\tilde X_i>c_{S_0}(p))}-1\bigg)\leq 0
\end{align*}
\end{enumerate}
Here, $\hat X_i= \bar \Phi^{-1}(\hat T_i)$, $\tilde X_i= \bar \Phi^{-1}(\tilde T_i)$ and $c_{S_0}(p)$ is the value s.t. $\P[\sum_{i\in S_0}\tilde X_i>c_{S_0}(p)]=p$.
\end{lem}

\mathleft
\begin{lem}\label{lem::mcond}
Let $\tilde\Omega_0=\{i: \tilde T_i \text{ follows } \Unif(0,1) \}$, $\mB_{0a}^{(\ell)}:=\{S\in\mB_0^\el:\exists A\in\mA^{(L)}\setminus \mA', s.t. S\subset A\}$,
and  $\mB_{0b}^{(\ell)}:=\{S\in\mB_{0a}^\el:S\in \tilde\Omega_0\}$,
we have:
\begin{flalign*}
\text{(1)}&\quad \max_{S\in\mB_{0a}^{(\ell)}}\sup_{c\in[0,\gamma_m]}\bigg|\frac{G_S(c)}{\bar\Phi(c)}-1\bigg|\to 0\\
\text{(2)}&
\quad \max_{S\in\mB_{0b}^{(\ell)}}\sup_{c\in[0,\bar\Phi^{-1}(1/m)]}\bigg|\frac{\P(X_S>c|\mQ^{(1:\ell-1)})}{\P(X_S>c)}-1\bigg|\to 0
\end{flalign*}
\end{lem}
\mathcenter

\begin{lem}
\label{lem::tail}
Define 
\begin{align}
&\mX^{(\ell)}=\Bigg\{
x:\sum_{S\in\mB_0^{(\ell)}}\csil I(T_S<\hat t^{(\ell)})-\sum_{S\in\mB_0^{(\ell)}}\csil \hat t^{(\ell)}
\leq 
\bigg\{\sum_{S\in\mB_0^{(\ell)}}\csil \hat t^{(\ell)}\bigg\}\epsilon
\Bigg\}\label{defx}\\
&\mX^{'(\ell)}=\Bigg\{
x:\bigg|\frac{\sum_{S\in\mB_0^{(\ell)}}\csil I(T_S<\hat t^{(\ell)})}{\sum_{S\in\mB_0^{(\ell)}}\csil \hat t^{(\ell)}}-1\bigg|
\geq 
\epsilon
\Bigg\}\nonumber
\end{align}
Then, $\forall \ell=1,...,L$, when the FDR control holds on layer $1,...,\ell-1$,
\begin{enumerate}[label=(\arabic*)]
\item For all $\epsilon\in(0,\alpha)$, if $\P(m\hat t^\el\geq Cc_\md)\to 1$, then $\P(\mX^{(\ell)})=1-o(1)$. Together with $\lim_{m\to\infty}|\tilde\Omega_0|/m=1$, we have $P(\mX^{'(\ell)})=1-o(1)$.
\item On $\cap_{h=1}^{\ell} \mX^{(h)}$, there exist a constant $C$ s.t. 
$\hat t^{(\ell)}\leq Cm^{r_1-1}$.
\item Let $\hat c_S$ be the rejection threshold for the test node $S\in\mB^{(\ell)}$, s.t. $\bar{G}_S(\hat c_S)=\hat t^{(\ell)}$. Then on
$\cap_{h=1}^{\ell} \mX^{(h)}$, 
\begin{equation*}
\hat c_S>\beta_m,\text{ }\forall S\in \mB^{(\ell)},
\end{equation*}
and on $\cap_{h=1}^{\ell-1} \mX^{(h)}$, 
\begin{equation*}
\hat c_S<\gamma_m,\text{ }\forall S\in \mB^{(\ell)}.
\end{equation*}
\end{enumerate}
\end{lem}

\begin{lem}\label{lem::equ}
\begin{equation}
\label{eq::lem4}
 \frac{\sumnull\csil\hat t ^{(\ell)}}{\sumall\csil I(T_S<\hat t^{(\ell)})}=\alpha(1+o(1))
 \end{equation}
\end{lem}
\vspace{1cm}


\begin{proof}[\textbf{Proof of Theorem~\ref{thm::layerFDR}}] Since the proof of the theorem statement (2) is similar to the proof of the theorem statement (1), we will only focusing on the proof of statement (1).\par
The random variable $FDP^{(\ell)}$ can be decomposed to the product of two parts.
\begin{align}
\label{FDPparts}
FDP^{(\ell)}=&\frac{\sumnull\csil I\{T_S<\hat t^{(\ell)}\}}{\sumnull\csil \hat t^{(\ell)}}\times \frac{\sumnull\csil \hat t^{(\ell)}}{\max(\sum_{S\in \mB^{(\ell)}}|S|I\{T_S<\hat t^{(\ell)}\},1)}
\end{align}
Based on \eqref{FDPparts}, in order to prove
$\lim_{m\to\infty}\P(FDP^{(\ell)}\leq \alpha+\epsilon)=1$
for all $\epsilon>0$, we only need prove
\begin{align}
&\lim_{m\to\infty}\P\Bigg\{\frac{\sumnull\csil I\{T_S<\hat t^{(\ell)}\}}{\sumnull\csil \hat t^{(\ell)}}-1<\epsilon\Bigg\}\to 1\label{part1}\\
&\lim_{m\to\infty}\P\Bigg\{\bigg|\frac{\sumnull\csil \hat t^{(\ell)}}{\max(\sum_{S\in \mB^{(\ell)}}|S|I\{T_S<\hat t^{(\ell)}\},1)}-\alpha\bigg|>\epsilon\Bigg\}\to 0\label{part2}
\end{align}
\eqref{part2} is immediately followed by Lemma \ref{lem::equ}, and we will prove \eqref{part1} by induction. Below is a list of the proof sketch:\par
\begin{enumerate}
\item On layer $1$, show $P(m\hat t^{(1)}\geq Cc_\md)\to 1$. Then, by applying Lemma~\ref{lem::tail}, we have 
\begin{itemize}
\item $P(\mX^{(1)})\to 1$, which is equivalent to \eqref{part1}. Hence, we proved the FDR control on layer 1.
\item $P(\beta_m<\hat c_S<\gamma_m,\forall S\in\mB^{(1)})\to 1$, and  $P(\hat c_S<\gamma_m,\forall S\in\mB^{(2)})\to 1$. Note that although this conclusion is not used to prove the FDR control on the current layer, but is necessary to guarantee the FDR control on higher layers.
\end{itemize}
\item On layer $\ell\geq 2$, assume the FDR control holds on previous layers and $P(\mX^{(\ell')})\to 1$ for all $\ell'=1,\ldots,\ell-1$. Then by Lemma~\ref{lem::tail}, $P(\beta_m<\hat c_S<\gamma_m,\forall S\in\cup_{\ell'=1}^{\ell-1}\mB^{(\ell')})\to 1$, and $P(\hat c_S<\gamma_m,\forall S\in \mB^\el)\to 1$. Accordingly, we can get
$P(m\hat t^{(1)}\geq  Cc_\md)\to 1$. Then, by applying the Lemma~\ref{lem::tail} again, we have
\begin{itemize}
\item $P(\mX^{(\ell)})\to 1$, which is equivalent to \eqref{part1}. Hence, we proved the FDR control on layer $\ell$.
\item  $P(\beta_m<\hat c_S<\gamma_m,\forall S\in\mB^{(\ell)})\to 1$, and $P(\hat c_S<\gamma_m,\forall S\in\mB^{(\ell+1)})\to 1$. 
\end{itemize}
\end{enumerate}

\par
We start the proof on layer 1.\par
\textbf{Layer 1:}\par
Take a subset $\mF^{(1)}\subset \mA_\md\cap\mA^{(1)}$, such that $|\mF^{(1)}|=c_\md$. 
For any $i\in \mF^{(1)}$, we have
$
\P(X_i>\gamma_m)\geq C
$.
By Markov's inequality, we have:
\begin{align*}
&\P\big(\big|\sum_{i\in \mF^{(1)}}I(X_i>\gamma_m)-\sum_{i\in\mF^{(1)}} \P(X_i>\gamma_m)|\geq c_\md^{3/4}\big)
\leq C(c_\md)^{-1/2}
\end{align*}

Thus, 
\begin{align*}
&\P\big[\sum_{1\leq i \leq m} I(T_i\leq \hat t^{(1)})\geq Cc_\md-c_\md^{3/4}\big]
\geq 1-o(1)
\end{align*}
Therefore, by Lemma \ref{lem::equ}, exists constant $C^{(1)}$, s.t.
\begin{align}
&\P\big[m_0\hat t^{(1)}\geq C^{(1)}c_\md\big]\geq 1-o(1)\label{mp1}
\end{align}
Together with Lemma \ref{lem::tail} (1), we have $\P(\mX^{(1)})\to 1$ and accordingly, $\P(FDP^{(1)}<\alpha+\epsilon)\to 1$.
\par
\vspace{0.5cm}
\textbf{Layer $\ell$:}\par
Based on similar arguments on Layer 1, it is suffice to show 
$\P(m_0\hat t^{(\ell)}>C^\el c_\md)\to 1$ for some constant $C^\el$.\par
Assume $\forall h=1,\ldots,\ell-1$, $\P(\mX^{(h)})\to 1$, then by Lemma \ref{lem::tail}, we have $\P(\beta_m<\hat c_S<\gamma_m,\forall S\in \mB^{(h)})\to 1$, and $\P( c_S<\gamma_m,\forall S\in \mB^{(\ell)})\to 1$.
\par
Let $\mF^{(\ell)}\subset  \mA_\md\cap\mA^{(\ell)}$ with $|\mF^{(\ell)}|=c_\md$. Define
\begin{equation*}
\hat\mF^{(\ell)}=\big\{A\in\mB^\el\cap \mF^\el:T_A<\alpha_m\big\}
\end{equation*}
By condition \ref{cond::largesignal}, $\forall A\in\mF^\el$,
\begin{align}
\P(A\in\hat \mF^\el)
\geq \P\big(T_A <\alpha_m,T_D \geq \bar\Phi(m^{r_1-1}\sqrt{\log m}), \forall D\in \mD(A)\big)
\geq & C_1\label{eql-0}
\end{align}
Accordingly, define
$\hat\mX^\el=\{|\hat \mF^{(\ell)}|\geq c_\md/2\}$,
then $\P(\hat\mX^\el)\geq 1-o(1)$. \par
On $\hat\mX^\el$, we have
\begin{equation*}
\sum_{S\in\mB_1^{(\ell)}}I(T_S\leq \hat t^{(\ell)})\geq Cc_\md
\end{equation*}
Then based on Lemma \ref{lem::mcond}, we can conclude that 
$\P(m_0\hat t^{(\ell)}\geq C^{(\ell)}c_\md)\geq 1-o(1)$
for some constant $C^{(\ell)}$.
\end{proof}

\begin{proof}[\textbf{Proof of Theorem~\ref{thm::overallFDR}}]
Let
$\mV^{(\ell)}=\{S\in\mB_0^{(\ell)}:S\subset\mR^{(\ell)}\}$ and
$\mW^{(\ell)}=\{S\in\mB_1^{(\ell)}:S\subset\mR^{(\ell)}\}$ be the
false rejection node set and the rejection node set on
layer $\ell$, respectively. 
Define
\begin{align*}
&\mX_1=\{S\in \cup_{\ell=2}^L \mW^{(\ell)}: S\cap \Omega_{\st}^{(1:L)}\neq\emptyset \text{ and }S\cap \Omega_0\neq \emptyset\}\\
&\mX_2=\{S\in \cup_{\ell=2}^L \mW^{(\ell)}: S\cap\Omega_{\wk}\neq \emptyset,\text{ } S\setminus(\Omega_{0}\cup\Omega_{\wk})=\emptyset\text{ and }S\cap \Omega_0\neq\emptyset\}\\
&\mX_3=\{S\in \cup_{\ell=2}^L \mW^{(\ell)}: S\cap \Omega_{1}\setminus(\Omega_{\wk}\cup\Omega_{\st}^{(1:L)})\neq \emptyset\text{ and }S\cap \Omega_0\neq\emptyset\}
\end{align*}

Then,
\begin{align*}
\P(\mX_1\neq \emptyset)&\leq \P(\mX_1\neq \emptyset|\cap_{\ell=1}^L\mX^\el)\P(\cap_{\ell=1}^L\mX^\el)+\P((\cap_{\ell=1}^L\mX^\el)^c)\leq Cm^{r_1}o(m^{-r_1})+o(1)\to 0\\
\P(\mX_2\neq \emptyset)
&\leq \P(\mX_2\neq \emptyset|\cap_{\ell=1}^L\mX^\el)\P(\cap_{\ell=1}^L\mX^\el)+\P((\cap_{\ell=1}^L\mX^\el)^c)\\
&\overset{(a)}{\leq} Cm^{r_1}\P\Bigg[X_S\geq \beta_m\bigg|S\in\Omega_\wk\cup\Omega_0\Bigg]+o(1)\\
&{\leq} Cm^{r_1}o(m^{-r_1})+o(1) \to 0
\end{align*}
Here, the inequality (a) is based on Lemma \ref{lem::tail} (1) and (3). By condition~\ref{cond::lip}, $|\mX_3|=o(c_\md)$, accordingly,
\begin{align*}
\P\big(FDP>\alpha+\epsilon\big)\leq&
\P(\mX_1\cup\mX_2\neq \emptyset)+\P\Bigg(\frac{\sum_{\ell=1}^L\sum_{S\in \mV^{(\ell)}} \csil}{\sum_{\ell=1}^L\sum_{S\in \mR_{\mathrm{node}}^{(\ell)}} \csil}>\alpha+\epsilon,\mX_1\cup\mX_2= \emptyset\Bigg)\\
\leq &o(1)+\sum_{\ell=1}^L \P\Bigg(\frac{\sum_{S\in \mV^{(\ell)}\setminus \mX_3}\csil}{\sum_{S\in \mR_{\mathrm{node}}^{(\ell)}}\csil}>\alpha+\epsilon+o(1)\Bigg)
\to 0
\end{align*}
So statement (1) is proved. The statement (2) can be proved in the similar way.
\end{proof}


\section*{Acknowledgment}

The microbiome samples were collected and sequenced at Memorial Sloan
Kettering Cancer Center (MSKCC) and pre-processed at Duke Cancer
Institute (DCI) Bioinformatics Shared Resource (BSR).  We thank Tsoni
Peled and Marcel van den Brink from MSKCC for their help in sample
collection and sequencing. We thank Kouros Owzar and Alexander Sibley
from DCI-BSR for the help in data pre-processing and constructive
discussions. Xuechan Li and Jichun Xie's research is supported by
Jichun Xie's startup fund from Duke University. Anthony Sung's
research is supported by NIH Award 1-R01-HL151365.



\bibliographystyle{elsarticle-harv}

\bibliography{manu.bib}

\newpage



\setcounter{section}{0}
\setcounter{subsection}{0}
\setcounter{equation}{0}
\setcounter{figure}{0}
\setcounter{table}{0}
\setcounter{page}{1}
\renewcommand\theequation{\arabic{equation}}


\renewcommand{\thepage}{S\arabic{page}}  
\renewcommand{\thesection}{S\arabic{section}}
\renewcommand{\thesubsection}{\thesection.\arabic{subsection}}
\renewcommand{\thetable}{S\arabic{table}}   
\renewcommand{\thefigure}{S\arabic{figure}}
\renewcommand{\theequation}{S\arabic{equation}}

%

 
 
\renewcommand{\bibnumfmt}[1]{[S#1]}
\renewcommand{\citenumfont}[1]{\textit{S#1}}
\renewcommand{\figurename}{Figure}



\begin{center}
{\large
\textbf{Supplementary Materials for 
  "Distance Assisted Recursive Testing"}
}
\end{center}

In this supplementary files, we provided the detailed algorithms of DART, the tuning parameter selection rule, numerical evaluation of setting $M$ as infinity, and the proofs of the lemmas.

\section{Algorithm Pseudo Codes} \label{sec::alg_code}

\subsection{Stage I: Transform the distance matrix into an aggregation tree} \label{sec::alg-stgi}

\begin{algorithm}[h] 
  \DontPrintSemicolon

  \KwData{distance matrix $\vD=(d_{ij})_{m\times m}$, the maximum
    layer $L$, the maximum children number $M$, the maximum distance
    threshold $g^{(2)},\ldots, g^{(L)}$.}

  \KwResult{an aggregation tree
    $\mT_L = \{\mA^\el: \ell \in\{1,\ldots, L\}\}$ and $\mC(A)$ for
    all $A\in \mA^\el$ and all $\ell\in\{1,\ldots,L\}$.}

  $\ell=1$, $\mA^{(1)}=\{\{1\},\ldots,\{m\}\}$,

  \lFor(\tcp*[h]{Remark~\ref{rem:l1-setup}}){$A\in
    \mA^{(1)}$}{$\mC(A)=\emptyset$ }
  
  \For(\tcp*[h]{Remark~\ref{rem:l-agg}}){$\ell\in \{2,\ldots, L\}$}
      { $\mA^\el=\emptyset$, $\tilde{\mA} = \mA^{(\ell-1)}$,   $\dist^\el(A_1,A_2)=\dist(A_1,A_2), \forall A_1,A_2\in \tilde\mA$
        \tcp*[h]{Remark~\ref{rem:Atilde}}\;

        \While{$|\tilde{\mA}\setminus \mA^\el|>0$}{
          $(\breve A_1,\breve A_2)=\underset{
            A_1\in\tilde{\mA},
            \ A_2\in\tilde{\mA}\setminus \{A_1\}
         }{\arg\min}\dist^\el(A_1, A_2)$ \tcp*[h]{Remark~\ref{rem:greedy}}\;
          
          \eIf(\tcp*[h]{Remark~\ref{rem:dia-thres}})
          {$\dist^\el(\breve A_1,\breve A_2) > g^{\el}$}
          {
              \lFor{$A \in \tilde{\mA}\setminus \mA^\el$}
               { $\mC(A) = A$, $\mA^\el = \mA^\el \cup \{A\}$, $\tilde{\mA}=\tilde\mA\setminus \{A\}$}
          }
          {$\breve A=\breve A_1\cup\breve A_2$ \;
               
            \For(\tcp*[h]{Remark~\ref{rem:set-children}})
            {$i\in \{1,2\}$} { \leIf{$\breve A_i\in \mA^\el$}
              { $\mC_i= \mC(\breve A_i)$}
              {$\mC_i = \{\breve A_i\}$} }
            $\mC(\breve A) = \mC_1 \cup \mC_2$\;
            \uIf{$|\mC(\breve A)| < M$} {
              $\mA^\el = \mA^\el \cup \{\breve A\} \setminus \{\breve
              A_1, \breve A_2\} $,
              $\tilde{\mA} = \tilde{\mA} \cup \{\breve A\} \setminus
              \{\breve A_1, \breve A_2\}$ }
              
              \uElseIf{$|\mC(\breve A)| = M$} {
              
              $\mA^\el = \mA^\el \cup \{\breve A\} \setminus \{\breve
              A_1, \breve A_2\} $,
              $\tilde{\mA} = \tilde{\mA}\setminus
              \{\breve A_1, \breve A_2\}$ \;} \uElse{
              $\dist^\el(\breve A_1,\breve A_2)=+\infty$
              \tcp*[h]{Remark~\ref{rem:check-children}}
              
            }
          }
        }
      }
      \caption{DART Stage I. Transform the distance matrix into an
        aggregation tree.}\label{alg:tree}
\end{algorithm}

  To obtain such an aggregation tree, we develop
  Algorithm~\ref{alg:tree} with remarks listed below.

\begin{remark}\label{rem:l1-setup}
  On layer 1, we set up each node as a single feature node. All these
  nodes have empty children sets.
\end{remark}

\begin{remark}\label{rem:l-agg}
  On layer $\ell$, we aggregate nodes from layer $\ell-1$ to form new
  nodes on this layer.
\end{remark}

\begin{remark}\label{rem:Atilde}
  At the beginning of layer $\ell$, $A^\el$ is set as the empty set,
  and it will be updated during the aggregation process.
  $\tilde{\mA}$ is the candidate node set with all the nodes that can
  possibly be aggregated. It may contain the layer $\ell-1$'s nodes
  that have not be aggregated yet and layer $\ell$'s nodes that have
  already been aggregated but can possibly be further aggregated. The layer $\ell$ distance between $A_1,A_2\in\tilde\mA$ is denoted by $\dist^\el(A_1,A_2)$. We set it equals to $\dist(A_1,A_2)$, which is defined in section \ref{sec::stgi}.
\end{remark}

\begin{remark}\label{rem:greedy}
  We use the greedy algorithm to select the closest two nodes
  $\breve A_1$ and $\breve A_2$ from the current candidate node set
  $\tilde{\mA}$. If there exists a tie, we select the first node pair that
  reaches the minimal distance. For example, in
  Figure~\ref{fig:dart-stgs}b, at the beginning of layer 2,
  $\dist(\{1\},\{2\})=2$ reaches the minimal distance among all node
  pairs on layer 1, so they will be selected to be further considered
  for aggregation.
\end{remark}

\begin{remark}\label{rem:dia-thres}
  We check if $\dist(\breve A_1, \breve A_2) > g^\el$. If yes, the
  remaining candidate nodes are too far away from each other and will
  not be further aggregated. Then the remaining child nodes on layer
  $\ell-1$ will be kept on layer $\ell$, and the aggregation on layer
  $\ell$ ends. If not, $\breve{A}_1$ and $\breve{A}_2$ will be further
  considered for aggregation.
\end{remark}

\begin{remark}\label{rem:set-children}
  We define the new node $\breve A = \breve A_1 \cup \breve
  A_2$. $\mC(\breve A)$ depends on the identity of $\breve A_1$ and
  $\breve A_2$: if $\breve A_i$ is a candidate child on layer
  $\ell-1$, then itself will be included in $\mC(\breve A)$; otherwise,
  $\breve A_i$'s children will be included in $\mC(\breve A)$.
\end{remark}

\begin{remark}\label{rem:check-children}
  We check the number of children of $\breve A$. If $|\mC(\breve A)|< M$, we add $\breve A$ to
  $\mA^\el$ and remove $\breve A_1$ and $\breve A_2$, and change
  $\tilde{\mA}$ correspondingly. If $|\mC(\breve A)|= M$, we change $\mA^\el$ in the same way when $|\mC(\breve A)|< M$, and remove $\breve A_1$ and $\breve A_2$ from $\tilde{\mA}$
  to prevent them being selected again. This step also guarantees the number of children of a node $A\in \tilde\mA$ is always smaller than $M$.  If $|\mC(\breve A)|> M$, we just reset the layer $\ell$ distance between $\breve A_1$ and $\breve A_2$ to be $+\infty$, so that they will never be aggregated on layer $\ell$, but still have chance to aggregate with other nodes in $\tilde\mA$.
\end{remark}

\newpage
\subsection{Stage II: Embed multiple testing in the tree}\label{sec::alg-stgii}

\begin{algorithm}[!htbp]
  \DontPrintSemicolon

  \KwData{Tree $\mT_L=\{\mA^{(i)}: i=1,\ldots,L\}$, P-values $(T_1,\ldots, T_m)$,
    and FDR level $\alpha$.}

  \KwResult{The set of rejected features $R_{\feat}$.}
    Set $\hat t^{(1)}$ as in \eqref{eq:t1}, $R_{\feat} = \{i: T_i
    \leq \hat t^{(1)}\}$ \tcp*[h]{Multiple testing on layer 1.}\; 

    \For(\tcp*[h]{Testing recursively on higher layers})
    {$\ell\in \{2,\ldots,L\}$}{
      $k = 0$, $T=$NULL\;
      \For{$S_k\in \mB^\el$}{ 
          $k=k+1$, $X_{S_k} = \sum_{j\in S_k} \bar{\Phi}^{-1}(T_j)/\sqrt{|S_k|}$\;
          $T = (T, \bar{\Phi}(X_{S_k}))'$ \tcp*[h]{Append the new working P-value at the end} 
      }
      Set $\hat{t}^{\el}$ as in \eqref{eq:tl},
      $R_{\node}^\el = \{S_{k'}: T_{k'} < \hat{t}^\el \}$,
      $R_{\feat} = R_{\feat} \cup \{\cup_{S\in R_{\node}^\el} S\}$
    }
 \caption{DART Stage II. Embed multiple testing in the tree.}\label{alg:test}
\end{algorithm}

\section{Numerical experiments}

\subsection{Simulated Settings} \label{sec::simu_settings}

Before we display the five settings, we first introduce the following
notations that are used across all five settings:
\begin{align*}
\eta_{1,i} & =\{[2\phi_1(d_{22,i})-0.2]\vee 0\}+\{\phi_2(d_{7,i})\};\\
\eta_{2,i} & =\{[3.4\phi_3(d_{156,i})-0.8]\vee 0\}+3\{\phi_4(d_{7,i})\}\\
&+10*I(i\in\{100,200,300,400,500,600,700,800,900,1000\}),
\end{align*}
where $\phi_1$, $\phi_2$, $\phi_3$ and $\phi_4$ are the PDF of
$\N(0,1)$, $\N(0,0.1)$,$\N(0,0.8)$ and $\N(0,0.05)$, respectively.

\begin{enumerate}[label=SE{\arabic*}:,align = left,leftmargin=*]
\item For node $i\in \{1,...,m\}$, the feature P-value
  $T_i=2\bar\Phi(|\breve Z_i|)$, where the $\breve Z_1,\ldots,\breve Z_m$
  are independently generated from $\N(\sqrt{n}\theta_i,1)$, with
$$\theta_{i}=\left\{
\begin{array}{ll}
\frac{1}{2}\eta_{1,i} I(\eta_{1,i}-0.15>0),& (n,m)=(90,100)\\
\frac{1}{7}\eta_{2,i} I(\eta_{2,i}-0.15>0),& (n,m)=(300,1000).
\end{array}
\right.$$

\item For node $i\in\{1,...,m\}$, the feature P-value
  $T_i=2\bar \Phi(|\breve Z_i|)$,
  where the $\breve Z_1,\ldots,\breve Z_m$
  are independently generated from a mixture distribution
  $0.04\mathrm{Laplace}(\sqrt{n}\theta_i,1)+0.96\mathrm{N}(\sqrt{n}\theta_i,1)$ with
$$\theta_{i}=\left\{
\begin{array}{ll}
\frac{2}{5}\eta_{1,i} I(\eta_{1,i}-0.15>0),& (n,m)=(90,100)\\
\frac{2}{13}\eta_{2,i} I(\eta_{2,i}-0.15>0),& (n,m)=(300,1000)
\end{array}
\right.$$
\item For node $i\in\{1,...,m\}$, the feature P-value
  $T_i=2\bar \Phi(|\breve Z_i|)$, 
 where the $\breve Z_1,\ldots,\breve Z_m$
  are independently generated from the mixture distribution with $0.04\mathrm t_{5}(\sqrt n\theta_i)+0.95\mathrm N(\sqrt n\theta_i,1)$. Here, $\mathrm t_{5}(\sqrt n\theta_i)$ stands for the student t distribution with $5$ degree of freedom and none centrality
  parameter $\sqrt{n}\theta_i$, with
$$\theta_{i}=\left\{
\begin{array}{ll}
\frac{1}{3}\eta_{1,i} I(\eta_{1,i}-0.15>0),& (n,m)=(90,100)\\
\frac{2}{13}\eta_{2,i} I(\eta_{2,i}-0.15>0),& (n,m)=(300,1000)
\end{array}
\right.$$
\item Consider the linear mode defined in \eqref{eq::lm}, with
  $p_0=3$, and $\sigma=1$. In model \eqref{eq::lm}, $W_1=1$ is the
  intercept term, $W_{2}$ and $W_{3}$ are sampled from
  $\mathrm{Binom}(0.5)$ and $\mathrm{Unif}(0.1,0.5)$, respectively.
  Also let $\theta_{1,i}=\theta_{3,i}=0.1$ and
$$\theta_{i}=\theta_{1,i}=\left\{
\begin{array}{ll}2\eta_{1,i} I(\eta_{1,i}-0.15>0),& (n,m)=(90,100)\\
\frac{5}{6}\eta_{2,i} I(\eta_{2,i}-0.15>0),& (n,m)=(300,1000)
\end{array}
\right.$$
 The feature P-value $T_i$ is defined in \eqref{eq::lmteststa}.
\item Consider the cox regression model
\begin{equation*}
\lambda_i(t)=\lambda_{0i}(t)\exp\{\theta_{1,i}W_{1}+\theta_{2,i}W_{2}\}
\end{equation*}
Where $\lambda_i(t)$ and $\lambda_{0i}(t)$ is the hazard and baseline hazard at time $t$, respectively. Set $\theta_{0,i}=\theta_{2,i}=0.1$ and
$$\theta_i=\theta_{1,i}=\left\{
\begin{array}{ll}\frac{4}{5}
\eta_{1,i} I(\eta_{1,i}-0.15>0),& (n,m)=(90,100)\\
\frac{2}{7}\eta_{2,i} I(\eta_{2,i}-0.15>0),& (n,m)=(300,1000)
\end{array}
\right.$$ The covariates $W_{1}$ and $W_{2}$ are sampled from
$\mathrm{Binom}(0.5)$ and $\mathrm{Unif}(0.1,0.5)$, respectively. The
event time is generated from the exponential distribution with rate
$\exp\{\theta_{1,i}W_{1}+\theta_{2,i}W_{2}\}$, and the censoring time
is sampled from $\Unif(0,5)$. The feature P-value $T_i$ is obtained
from the Wald test.
\end{enumerate}

\subsection{Tuning parameter selection for applying DART on simulated data}
\label{sec::simu_tuning}
The section 2.3 introduces the tuning parameter selection for the aggregation tree construction. Based on it, the tuning parameter for our numerical study is selected as follow:
\begin{itemize}
\item If $(n,m)=(90,100)$: Based on recommendation in section 2.3, we choose $M=3$ and construct a $L=\lceil\log_M 100-\log_M 30\rceil=2$ layers aggregation tree.
We use Algorithm~\ref{alg:g} to construct a dynamic set $G$ and search the value $g^{(2)}$. Table~\ref{tab::m3} (1) tracks the number of non-single-child nodes $|\tilde A^{(2)}(g)|$ based on different values of $g\in G$. By applying the algorithm, 
 $$g^{(2)}=26/\sqrt{n\log m\log\log m}$$
\item If $(n,m)=(300,1000)$: Similar to the previous case, we choose $M=3$ and construct a $L=\lceil\log_M 1000-\log_M 30\rceil=4$ layers aggregation tree.
Based on Algorithm~\ref{alg:g}, we have Table~\ref{tab::m3} which tracks the number of non-single-child nodes on each layer, and,
\begin{align*}
g^{(2)}=\frac{8}{\sqrt{n\log m\log\log m}},
g^{(3)}=\frac{22}{\sqrt{n\log m\log\log m}},
g^{(4)}=\frac{56}{\sqrt{n\log m\log\log m}}
\end{align*}

\end{itemize}

\newpage
\begin{algorithm}[!htp]
  \KwData{Distance Matrix $D=(d_{ij})_{m\times m}$, Sample size $n$, number of features $m$, the maximum children number $M$, the maximum layer L}

  \KwResult{$g^{(2)},...,g^{(L)}$.}
  \tcp{set searching upper bound $d_{\max}$ and step-size $s_{n,m}$}
  Let $d_{\max}=\max_{j\in\Omega}\min_{i\in\{i:i\neq j\}}d_{ij}$; $s_{n,m}=2/\sqrt{n\log(m)\log\log(m)}$
  \;

  \For{$\ell=2,...,L$}{\tcp{on layer $\ell$, search $g^\el$ from $(g^{(\ell-1)},d_{\max}]$, $g^{(1)}=0$}
      Let $M_g=$NULL;  $e_g$=1;
      $G=$NULL; $g=g^{(\ell-1)}+s_{n,m}$\;
\While{$g\leq(2M^{L-2}-1)d_{\max}$ and $e_g<10$
}{\tcp{stop searching process if the value $g$ exceed the searching upper bound or the $|\tilde A^\el(g)|$ does not increase for past $10$ candidate values $g$.}
\tcp{stop searching process if the value $g$ exceed the searching upper bound.}
Use Algorithm \ref{alg:tree} to Construct an $\ell$ layers aggregation tree $\mT_{\ell}=\{\mA^{(\ell')}:\ell'=1,...,\ell\}$ with maximum children number M, and $(g^{(1)},..., g^{(\ell-1)},g)$\;
Set $\tilde A^\el(g) = \{A: A\in \mA^\el(g),\ |\mC(A)|\geq 2\}$; 
\If{$m_g\geq |\tilde A^\el(g)|$}{
$e_g=e_g+1$\;
}\Else{$e_g$=1\;}

$G=(G,g)$; $M_g=(M_g,|\tilde A^\el(g)|)$
; $m_g=|\tilde A^\el(g)|$
\;
$g=g+s_{n,m}$\;
}
$g^\el=\min\{\arg\max_{g\in G}M_g$\}\;
}

 \caption{$g^{(\ell)}$ Selection algorithm.}\label{alg:g}
\end{algorithm}

\begin{table}[!htp]
\centering
\caption{The number of non-single-child nodes based on value $g\in G$ when $M=3$. For simplicity purpose, the value $g$ is represented by its nominator: $g'=g\times \sqrt{n\log m\log\log m}$. The selected $g'$ and its correspnding $|\tilde\mA^{(2)}(g)|$ is highlighted in bold.}
\label{tab::m3}
\resizebox{\columnwidth}{!}{%
\begin{tabular}{l|c|rrrrrrrrrrrrrrrrrrrr}
\multicolumn{13}{l}{}\\
\multicolumn{13}{l}{(1) $(n,m)=(90,100)$:}\\
\cline{1-19}
\multirow{2}{*}{Layer 2}
&$g'$ & 2 & 4 & 6 & 8 & 10 & 12 & 14 & 16 & 18 & 20 & 22 & 24 & \textbf{26} & 28 & 30 & 32 & $\ldots$ \\ 
\cline{2-19}
 &$|\tilde\mA^{(3)}(g)|$ & 5 & 10 & 17 & 22 & 29 & 31 & 31 & 39 & 40 & 40 & 40 & 40 & \textbf{41} & 41 & 41 & 41 & $\ldots$\\ 
  \cline{1-19}
\multicolumn{13}{l}{}\\
\multicolumn{13}{l}{(2) $(n,m)=(300,1000)$:}\\
\hline
\multirow{2}{*}{Layer 2}&
$g'$ & 2     & 4     & 6     & \textbf{8}     & 10     & 12     & 14     & 16     &$\ldots$ & & & &\\ 
\cline{2-19}
&$|\tilde\mA^{(2)}(g)|$ & 49     & 149     & 245     & \textbf{293}     & 293     & 293     & 293     & 293  & $\ldots$ & & &  &   \\
\hline
\multirow{2}{*}{Layer 3}&$g'$& 10 & 12   & 14   & 16   & 18   & 20   & \textbf{22}   & 24   & 26   & 28   & 30 & $\ldots$\\ 
\cline{2-19}
 &$|\tilde\mA^{(3)}(g)|$& 103& 154   & 191   & 221   & 230   & 239   & \textbf{241}   & 241   & 241   & 241   & 241 & $\ldots$ \\ 
\hline
\multirow{2}{*}{Layer 4}&$g'$ &$\ldots$ & 38 & 40 & 42  & 44  & 46  & 48  & 50  & 52  & 54  & \textbf{56}  & 58  & 60  & 62  & 64 & $\ldots$ \\ 
\cline{2-19}
 &$|\tilde\mA^{(4)}(g)|$  & $\ldots$  & 116  & 118  & 119  & 120  & 120  & 120  & 120  & 120 & 120  & \textbf{121}  & 121  & 121  & 121  & 121 & $\ldots$\\ 
\hline
\end{tabular}%
}
\end{table}

\section{Additional numerical results for assessing impact of the parameter \texorpdfstring{$M$}{M}}
\label{sec::impactm}
In this section, we numerically investigate the impact of the choice of $M$ by comparing the numerical results when $M=3$ and  $M=\infty$. When $M=3$, the tunning parameters are same to the parameters in \ref{sec::tuning}. When $M=\infty$, in order to have a relatively fair comparison, we set the same total layer $L$ as the value in \ref{sec::tuning}. The selection procedure of $g^{(\ell)}$ is similar to \ref{sec::tuning}.  Based on the Algorithm~\ref{alg:g},we have
\begin{itemize}
\item If $(n,m)=(90,100)$:  we set $g^{(2)}=\frac{16}{\sqrt{n\log m\log\log m}}$
\item If $(n,m)=(300,1000)$:  we set
$g^{(2)}=\frac{12}{\sqrt{n\log m\log\log m}}$, $g^{(3)}=\frac{26}{\sqrt{n\log m\log\log m}}$ and $g^{(4)}=\frac{44}{\sqrt{n\log m\log\log m}}$.
\end{itemize} 

\begin{figure}[!ht]
\centering
\includegraphics[scale=0.085]{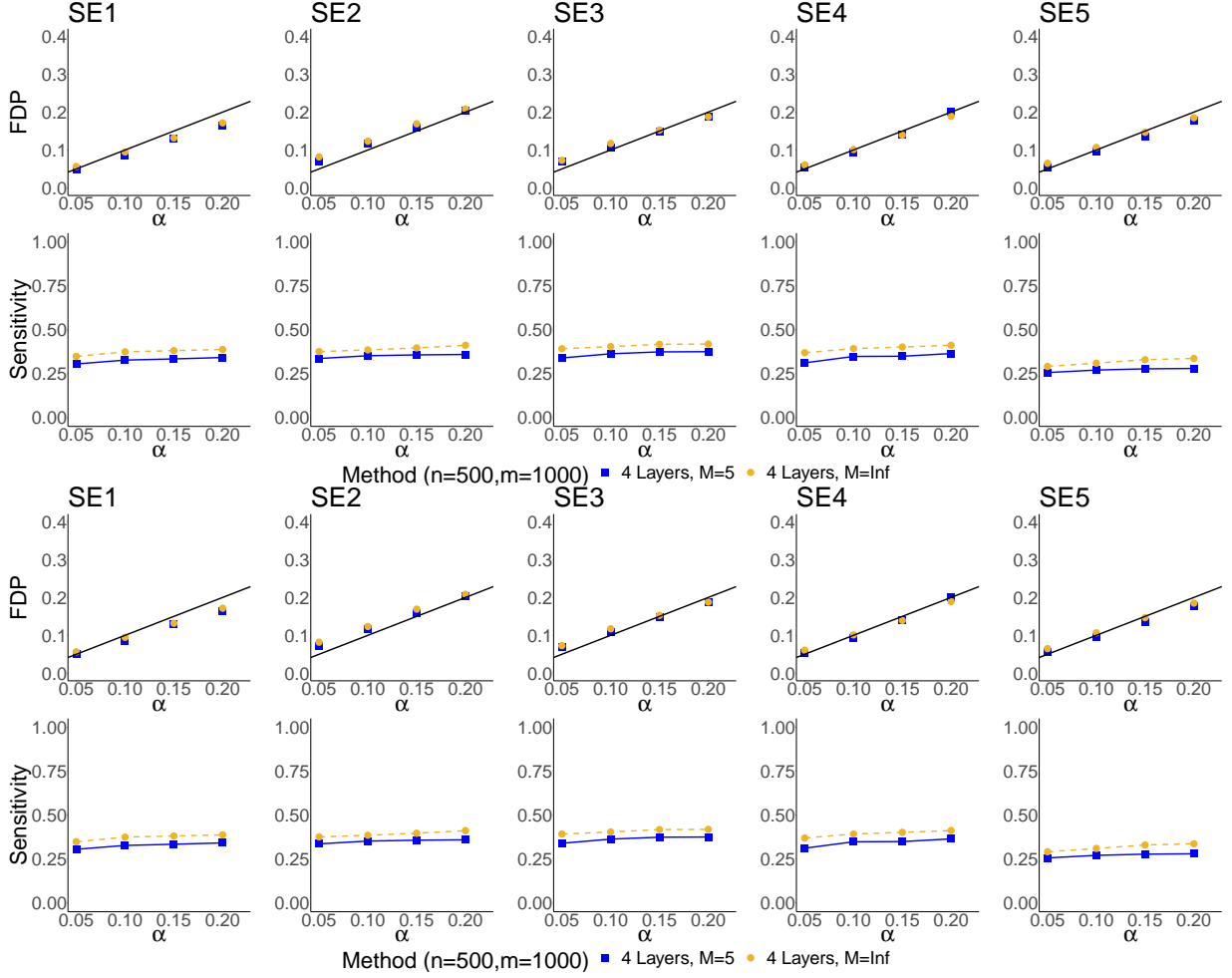}
\caption{\it Additional simulation results for setting SE1-SE5. The first two rows represent the results in the setting $(n,m)=(90,100)$, and the second two rows represent the results in the setting $(n,m)=(300,1000)$.}
\label{fig::addsim}
\end{figure}

Figure~\ref{fig::addsim} compares the performance between two different $M$ values under SE1-SE5. We only compare the performance on the top layer of the aggregation tree. Based on the figure, our method is still valid with FDR control when $M=\infty$.

\begin{table}[!ht]
\centering
\caption{The number of non-single-child nodes based on value $g\in G^\el$ when $M=\infty$. For simplicity purpose, the value $g$ is represented by its nominator: $g'=g\times \sqrt{n\log m\log\log m}$. The selected $g'$ and its correspnding $|\tilde\mA^\el(g)|$ is highlighted in bold.}
\label{tab::minf}
\resizebox{\columnwidth}{!}{%
\begin{tabular}{l|c|rrrrrrrrrrrrrrrrr}
\multicolumn{14}{l}{}\\
\multicolumn{14}{l}{(1) $(n,m)=(90,100)$:}\\
\cline{1-15}
  \multirow{2}{*}{Layer 2}&
$g'$ & 2 & 4 & 6 & 8 & 10 & 12 & 14 & \textbf{16} & 18 & 20 & 22 & 24 &$\ldots$\\ 
\cline{2-15}
 &$|\tilde\mA^{(2)}(g)|$ & 5 & 10 & 17 & 22 & 29 & 30 & 30 & \textbf{35} & 32 & 31  & 29 & 28 & $\ldots$\\ 
\cline{1-15}
\multicolumn{16}{l}{}\\
\multicolumn{16}{l}{(2) $(n,m)=(300,1000)$:}\\
  \hline
\multirow{2}{*}{Layer 2}&$g'$ & 2 & 4 & 6 & 8 & 10 & \textbf{12} & 14 & 16 & 18 & 20     &$\ldots$ & & & \\ 
\cline{2-16}
 &$|\tilde\mA^{(2)}(g)|$ & 49 & 149 & 239 & 291 & 300 & \textbf{303} & 295 & 288 & 263 & 245 &$\ldots$ & & & \\ 
   \hline
\multirow{2}{*}{Layer 3}&
$g'$ & 14 & 16 & 18 & 20 & 22 & 24 & \textbf{26} & 28 & 30 & 32 & 34 &$\ldots$\\ 
\cline{2-16}
 &$|\tilde\mA^{(3)}(g)|$ & 53 & 101 & 130 & 148 & 163 & 167 & \textbf{169} & 166 & 159 & 152 & 144&$\ldots$\\ 
   \hline

\multirow{2}{*}{Layer 4}&
$g'$ & 28 & 30 & 32 & 34 & 36 & 38 & 40 & 42 & \textbf{44} & 46 & 48 & 50 & 52 & $\ldots$\\ 
\cline{2-16}
 &$|\tilde\mA^{(4)}(g)|$& 17 & 33 & 47 & 56 & 66 & 70 & 74 & 76 & \textbf{80} & 79 & 79 & 79 & 77 & $\ldots$ \\ 
   \hline
\end{tabular}%
}
\end{table}

\section{Proof of the Lemmas}

\begin{proof}[Proof of Lemma \ref{lem:lin-reg-T}]
Let $X^{'\ast}_{i}=\frac{\pmb q^T\pmb{\hat\theta_i}}{s\sqrt{\pmb q^T(\pmb \mW^T\pmb \mW)^{-1}\pmb q}}$ and $X'_{o,i}=\frac{\pmb q^T\pmb{\hat\theta_i}}{\sigma\sqrt{\pmb q^T(\pmb \mW^T\pmb \mW)^{-1}\pmb q}}$, we have $X'_{o,i}\sim N(\eta_i,1)$ with $\eta_i=\frac{\pmb q^T\pmb \theta_i}{\sigma\sqrt{\pmb q^T(\pmb \mW^T\pmb \mW)^{-1}\pmb q}}$. 

To show the statistics $T_i$ is asymptotically oracle, it is suffice to show:
\begin{equation*}
P(|\Phi^{-1}(T_i)-\Phi^{-1}(\tilde T_{i})|>(\log m)^{-2.5})=o((\log m)^{-1})
\end{equation*}

Given
\begin{align*}
|\Phi^{-1}(T_i)-\Phi^{-1}(\tilde T_{i})|&\leq \sup_{x\geq 0}\frac{\phi(x)}{\phi[\Phi^{-1}(2\Phi(-x))]}||X_i^{'\ast}|-|X'_{o,i}||\\
&\leq |X_i^{'\ast}-X'_{o,i}|\\
&=|X'_{o,i}(\sigma/s-1)|
\end{align*}
and based on condition~\ref{cond::sparse},
\begin{align*}
P\bigg(|X'_{o,i}(\sigma/s-1)|>(\log m)^{-2.5}\bigg)\leq& P(|X'_{o,i}|>\sqrt{\log m})+P(|\sigma/s-1|>(\log m)^{-3})
\end{align*}

It is suffice to show
\begin{equation}
\label{eq::lm-approx}
P(|\sigma/s-1|>(\log m)^{-3})=o((\log m)^{-1})
\end{equation}

Let $Y_1,...,Y_{n-p_0-1}\overset{iid}{\sim} \mX^2(1)$ and $Y=\sum_{k=1}^{n-p_0-1}(Y_k-1)/\sqrt{2(n-p_0-1)}$, we have $Y/\sqrt{n-p_0-1}\sim \mX^2(n-p_0-1)/(n-p_0-1)-1$. Since $s^2/\sigma^2\sim \mX^2(n-p_0-1)/(n-p_0-1)$, based on Lemma 6.1 in \citet{liu2013gaussian},
\begin{align*}
P(|s/\sigma-1|>(\log m)^{-3})\leq & P(|s^2/\sigma^2-1|>(\log m)^{-3})=o((\log m)^{-1})
\end{align*}
Thus, after trivial calculation, the equation \eqref{eq::lm-approx} holds. 
\end{proof}

\begin{proof}[Proof of Lemma \ref{lem::asym}]
(1) Define $\tilde X_i=\bar\Phi(\tilde T_i)$, For $k\in\{1,...,c_0\}$, let
 $q_{0}\geq \epsilon(m)$.
Also define $b_{1,k}(q_0)$, $c_{1}$,...,$c_{k}$ be the value s.t. $P(\sum_{j=1}^k\tilde X_j>b_{1,k}(q_0))=q_0\big[\epsilon'(m)\big]^{(c_0-k)/c_0}$, and $P(\tilde X_1>c_{1})=...=P(\tilde X_k>c_{k})=\epsilon(m)\epsilon'(m)$, respectively. For simplicity sake, we use $b_{1,k}$ to present $b_{1,k}(q_0)$.
 
Based on the definition, we have
 \[b_{1,k}<\sum_{j=1}^k c_j\]
 \par
 Thus, when $k=2$,
\begin{align*}
&P(\hat X_1+\hat X_2>b_{1,2})\\
=& P(\hat X_1+\hat X_2>b_{1,2},\hat  X_1>b_{1,2}-c_2,\hat  X_2>b_{1,2}-c_1)\\ &+P(\hat X_1+\hat  X_2>b_{1,2},\hat  X_1<b_{1,2}-c_2)+P(\hat X_1+\hat  X_2>b_{1,2},\hat  X_2<b_{1,2}-c_1)\\
= & P(\hat X_1+\hat  X_2>b_{1,2},c_1>\hat  X_1>b_{1,2}-c_2)+P(\hat  X_1>c_1,\hat  X_2>b_{1,2}-c_1)\\ &+P(\hat  X_1+\hat  X_2>b_{1,2},\hat  X_1<b_{1,2}-c_2)+P(\hat  X_1+\hat  X_2>b_{1,2},\hat  X_2<b_{1,2}-c_1)
\end{align*}
Based on construction, the last three terms always smaller than $\epsilon(m)\epsilon'(m)(1+\delta_4(m))$ for $\delta_4(m):=\max_{i\in \Omega}\sup_{p\in \mP'_i}\bigg|P(\hat T_i<p)/P(\tilde T_i<p)-1\bigg|\to 0$, and accordingly, we have 
\begin{align*}
& P(\hat  X_1+\hat  X_2>b_{1,2},c_1>\hat  X_1>b_{1,2}-c_2)+P(\hat  X_1>c_1,\hat  X_2>b_{1,2}-c_1)\\ 
\leq & [P(\hat  X_1+\tilde X_2>b_{1,2},c_1>\hat  X_1>b_{1,2}-c_2)+P(\hat  X_1>c_1,\tilde X_2>b_{1,2}-c_1)](1+\delta_4(m))\\ 
\leq & [P(\hat X_1+\tilde X_2>b_{1,2},\hat  X_1>b_{1,2}-c_2,\tilde X_2>b_{1,2}-c_1)](1+\delta_4(m))\\
\leq & P(\tilde X_1+\tilde X_2>b_{1,2},\tilde  X_1>b_{1,2}-c_2,\tilde X_2>b_{1,2}-c_1)(1+\delta_4(m))^2
\end{align*}
Based on similar arguments, we can also have
\begin{align*}
& P(\hat  X_1+\hat  X_2>b_{1,2},c_1>\hat  X_1>b_{1,2}-c_2)+P(\hat  X_1>c_1,\hat  X_2>b_{1,2}-c_1)\\ 
\geq & P(\tilde X_1+\tilde X_2>b_{1,2},\tilde  X_1>b_{1,2}-c_2,\tilde X_2>b_{1,2}-c_1)(1-\delta_4(m))^2
\end{align*}
Thus,
\begin{equation*}
\sup_{q_0\geq \epsilon(m)
\big[\epsilon'(m)\big]^{\frac{c_0-2}{c_0}}}\bigg|\frac{P(\hat X_1+\hat X_2>b_{1,2})}{P(\tilde X_1+\tilde X_2>b_{1,2})}-1\bigg|\to 0
\end{equation*}
Similarly, if $\sup_{q_0\geq \epsilon(m)\big[\epsilon'(m)\big]^{\frac{c_0-k}{c_0}}}\bigg|\frac{P(\sum_{j=1}^k\hat X_j>b_{1,k})}{P(\sum_{j=1}^k\tilde X_j>b_{1,k})}-1\bigg|\to 0$, we can have
\begin{equation*}
\sup_{q_0\geq \epsilon(m)\big[\epsilon'(m)\big]^{\frac{c_0-k-1}{c_0}}}\bigg|\frac{P(\sum_{j=1}^{k+1}\hat X_j>b_{1,k+1})}{P(\sum_{j=1}^{k+1}\tilde X_j>b_{1,k+1})}-1\bigg|\to 0
\end{equation*}
Thus, we can get (1). In addition, based on the similar arguments, we can get (2).

\end{proof}

\begin{proof}[Proof of Lemma \ref{lem::mcond}]
(1) Let $Z'_1,...,Z'_K \overset{iid}{\sim} N(0,1)$, with $2 \leq K < M^{L-1}$. Define the set $\mathfrak{M}=\{\mM_1\subset \{1,...,m\}: 1 \leq |\mM_1|\leq K-1\}$. It is suffice to show:
\begin{equation*}
\lim_{m\to \infty}\sup_{\mM_1\in\mathfrak{M}}\sup_{\substack{c_1\in [\beta_0,\gamma_m] \\ c_2\in [0,
\gamma_m]
}} \frac{\P(\frac{1}{\sqrt K}\sum_{i=1}^K Z'_i>c_2,\frac{1}{\sqrt{|\mM_1|}}\sum_{j\in \mM_1}Z'_j> c_1)}{\P(\frac{1}{\sqrt K}\sum_{i=1}^K Z'_i>c_2)}
= 0
\end{equation*}
Here, $\beta_0=\sqrt{2b(1-r_1)\log m+b(1-r_1)\log\log\log m}$, with \[b=\frac{\frac{2M^{L-1}+1}{M^{L-1}+1}-r_1}{2(1-r_1)}\in\bigg(\frac{M^{L-1}}{(M^{L-1}+1)(1-r_1)},1\bigg).\]
For simplification, let $k_1=|\mM_1|$. For $Z_1$ and $Z_2\overset{iid}{\sim}\N(0,1)$, define  $$\mD_m=\bigg\{c_2\in (0,\gamma_m): \frac{d}{dc_2}\frac{P\big(\sqrt{\frac{k_1}{K}}Z_1+\sqrt{\frac{K-k_1}{K}}Z_2>c_2,Z_1>\beta_0\big)}{P\big(\sqrt{\frac{k_1}{K}}Z_1+\sqrt{\frac{K-k_1}{K}}Z_2>c_2\big)}=0\bigg\}$$
, then 
\begin{align*}
&
\sup_{\substack{c_1\in [\beta_0,\gamma_m] \\ c_2\in [0,
\gamma_m]
}} \frac{P(\frac{1}{\sqrt K}\sum_{i=1}^K Z'_i>c_2,\frac{1}{\sqrt{|\mM_1|}}\sum_{j\in \mM_1}Z'_j> c_1)}{P(\frac{1}{\sqrt K}\sum_{i=1}^K Z'_i>c_2)}\\
\leq&
2\sup_{ c_2\in [0,
\gamma_m]
}\frac{P\big(\sqrt{\frac{k_1}{K}}Z_1+\sqrt{\frac{K-k_1}{K}}Z_2>c_2,Z_1>\beta_0\big)}{P\big(\sqrt{\frac{k_1}{K}}Z_1+\sqrt{\frac{K-k_1}{K}}Z_2>c_2\big)}\\
\leq& 2 \max\Bigg\{\max_{c_2= 0\text{ or }  \gamma_m}\frac{P\big(\sqrt{\frac{k_1}{K}}Z_1+\sqrt{\frac{K-k_1}{K}}Z_2>c_2,Z_1>\beta_0\big)}{P\big(\sqrt{\frac{k_1}{K}}Z_1+\sqrt{\frac{K-k_1}{K}}Z_2>c_2\big)},\\
&\sup_{c_2\in \mD_m}\frac{P\big(\sqrt{\frac{k_1}{K}}Z_1+\sqrt{\frac{K-k_1}{K}}Z_2>c_2,Z_1>\beta_0\big)}{P\big(\sqrt{\frac{k_1}{K}}Z_1+\sqrt{\frac{K-k_1}{K}}Z_2>c_2\big)}\Bigg\}
\end{align*}
\par
(i). When $c_2=0$,
{\footnotesize
\[
\lim_{m\to \infty}\frac{P\big(\sqrt{\frac{k_1}{K}}Z_1+\sqrt{\frac{K-k_1}{K}}Z_2>c_2,Z_1>\beta_0\big)}{P\big(\sqrt{\frac{k_1}{K}}Z_1+\sqrt{\frac{K-k_1}{K}}Z_2>c_2\big)}\\
=\lim_{m\to \infty}2P\big(\sqrt{\frac{k_1}{K}}Z_1+\sqrt{\frac{K-k_1}{K}}Z_2>c_2,Z_1>\beta_0\big) =0
\]
}
(ii). When $c_2=\gamma_m$, $c_2/\beta_0=\sqrt{\frac{1}{b(1-r_1)}}$,
{\footnotesize
\begin{align*}
& \lim_{m\to \infty}\frac{P\big(\sqrt{\frac{k_1}{K}}Z_1+\sqrt{\frac{K-k_1}{K}}Z_2>c_2,Z_1>\beta_0\big)}{P\big(\sqrt{\frac{k_1}{K}}Z_1+\sqrt{\frac{K-k_1}{K}}Z_2>c_2\big)}
=\lim_{\beta_0\to\infty}\frac{\int_{\beta_0}^\infty\int^\infty_{S\sqrt{\frac{K}{K-k_1}}\beta_0-\sqrt{\frac{k_1}{K-k_1}}z_1}\phi(z_1)\phi(z_2)dz_2dz_1}{\int_{S\beta_0}^\infty \phi(z)dz}\\
\leq& C\lim_{\beta_0\to \infty}\frac{\int^\infty_{S\sqrt{\frac{K}{K-k_1}}\beta_0-\sqrt{\frac{k_1}{K-k_1}}\beta_0}\phi(\beta_0)\phi(z)dz+\int_{\beta_0}^\infty\phi(z)\phi(S\sqrt{\frac{K}{K-k_1}}\beta_0-\sqrt{\frac{k_1}{K-k_1}}z)dz}{\phi(S\beta_0)}\text{ (L'Hopital's rule)}\\
\leq & C\lim_{\beta_0\to\infty}\Bigg[\exp\bigg\{-\frac{\beta_0^2}{2}\bigg(S\sqrt{\frac{k_1}{K-k_1}}-\sqrt{\frac{K}{K-k_1}}\bigg)^2\bigg\}
+\int_{\beta_0}^\infty\exp\bigg\{-\frac{1}{2}\bigg(\sqrt{\frac{K}{K-k_1}}z-S\sqrt{\frac{k_1}{K-k_1}}\beta_0\bigg)^2\bigg\}dz\Bigg]
=0,
\end{align*}
}
Where $S=\sqrt{\frac{1}{b(1-r_1)}}$\par
(iii). When $c_2\in \mD_m$, given
\begin{align*}
0=&\frac{d}{dc_2}\frac{P\big(\sqrt{\frac{k_1}{K}}Z_1+\sqrt{\frac{K-k_1}{K}}Z_2>c_2,Z_1>\beta_0\big)}{P\big(\sqrt{\frac{k_1}{K}}Z_1+\sqrt{\frac{K-k_1}{K}}Z_2>c_2\big)}\\=&\frac{1}{P\big(\sqrt{\frac{k_1}{K}}Z_1+\sqrt{\frac{K-k_1}{K}}Z_2>c_2\big)^2} \times\\
&\bigg\{P\big(\sqrt{\frac{k_1}{K}}Z_1+\sqrt{\frac{K-k_1}{K}}Z_2>c_2\big)\frac{d}{dc_2}P\big(\sqrt{\frac{k_1}{K}}Z_1+\sqrt{\frac{K-k_1}{K}}Z_2>c_2,Z_1>\beta_0\big)\\
&-P\big(\sqrt{\frac{k_1}{K}}Z_1+\sqrt{\frac{K-k_1}{K}}Z_2>c_2,Z_1>\beta_0\big)\frac{d}{dc_2}P\big(\sqrt{\frac{k_1}{K}}Z_1+\sqrt{\frac{K-k_1}{K}}Z_2>c_2\big)\bigg\}
\end{align*}
We have
\begin{align*}
\frac{P\big(\sqrt{\frac{k_1}{K}}Z_1+\sqrt{\frac{K-k_1}{K}}Z_2>c_2,Z_1>\beta_0\big)}{P\big(\sqrt{\frac{k_1}{K}}Z_1+\sqrt{\frac{K-k_1}{K}}Z_2>c_2\big)}=\frac{\frac{d}{dc_2}P\big(\sqrt{\frac{k_1}{K}}Z_1+\sqrt{\frac{K-k_1}{K}}Z_2>c_2,Z_1>\beta_0\big)}{\frac{d}{dc_2}P\big(\sqrt{\frac{k_1}{K}}Z_1+\sqrt{\frac{K-k_1}{K}}Z_2>c_2\big)}
\end{align*}
Therefore,
\begin{align*}
&\sup_{c_2\in \mD_m}\frac{P\big(\sqrt{\frac{k_1}{K}}Z_1+\sqrt{\frac{K-k_1}{K}}Z_2>c_2,Z_1>\beta_0\big)}{P\big(\sqrt{\frac{k_1}{K}}Z_1+\sqrt{\frac{K-k_1}{K}}Z_2>c_2\big)}\\
=&\sup_{c_2\in \mD_m}\frac{\frac{d}{dc_2}P\big(\sqrt{\frac{k_1}{K}}Z_1+\sqrt{\frac{K-k_1}{K}}Z_2>c_2,Z_1>\beta_0\big)}{\frac{d}{dc_2}P\big(\sqrt{\frac{k_1}{K}}Z_1+\sqrt{\frac{K-k_1}{K}}Z_2>c_2\big)}\\
=&\sup_{c_2\in \mD_m}C\int_{\beta_0}^\infty \exp\bigg\{-\frac{1}{2}\bigg(\sqrt{\frac{K}{K-k_1}}z-\sqrt{\frac{k_1}{K-k_1}}c_2\bigg)^2\bigg\}dz\\
\leq &C\int_{\beta_0}^\infty \exp\bigg\{-\frac{1}{2}\bigg(\sqrt{\frac{K}{K-k_1}}z-\sqrt{\frac{k_1}{K-k_1}}\gamma_m\bigg)^2\bigg\}dz\\
\to & 0
\end{align*}
Combine (i), (ii) and (iii), we have 
\begin{equation*}
\lim_{m\to \infty}\sup_{\mM_1\in\mathfrak{M}}\sup_{\substack{c_1\in [\beta_0,\gamma_m] \\ c_2\in [0,
\gamma_m]
}} \frac{P(\frac{1}{\sqrt K}\sum_{i=1}^K Z_i>c_2|\frac{1}{\sqrt{|\mM_1|}}\sum_{j\in \mM_1}Z_j|> c_1)}{P(\frac{1}{\sqrt K}\sum_{i=1}^K Z_i>c_2)}
= 0
\end{equation*}
(2)\par
It is suffice to show 
\begin{equation*}
\lim_{m\to \infty}\sup_{\mM_1\in\mathfrak{M}}\sup_{\substack{ c_2\in [0,
\bar\Phi^{-1}(1/m)]
}} \frac{\P(\frac{1}{\sqrt K}\sum_{i=1}^K X_i>c_2,\frac{1}{\sqrt{|\mM_1|}}\sum_{j\in \mM_1}X_j> \beta_0)}{\P(\sum_{i=1}^KZ'_i/\sqrt{K}>c_2)}
\leq 0
\end{equation*}
Let $\breve X_1=\sum_{i\in\mM_1}X_i/\sqrt{k_1}$, $\breve X_2=\sum_{i\in\mathfrak{M}\setminus\mM_1}X_i/\sqrt{K-k_1}$.\par
Based on lemma~\ref{lem::asym}, $\delta_{6m}=|P(\breve X_j>p)/P( Z_j>p)-1|\to 0$ uniformly for $j=1,2$ and $p>\alpha_m$.
\par
Thus, uniformly,
\begin{align*}
&P(\sqrt{\frac{k_1}{K}}\breve X_1+\sqrt{\frac{K-k_1}{K}}\breve X_2>c_2,\breve X_1>\beta_0)\\
=& P(\sqrt{\frac{K-k_1}{K}}\breve X_2>c_2-\sqrt{\frac{k_1}{K}}\beta_0,\breve X_1>\beta_0)\\
&+P(\sqrt{\frac{K-k_1}{K}}\breve X_2<c_2-\sqrt{\frac{k_1}{K}}\beta_0,\sqrt{\frac{k_1}{K}}\breve X_1+\sqrt{\frac{K-k_1}{K}}\breve X_2>c_2)\\
\leq & (1+\delta_{6m})\big[P(\sqrt{\frac{K-k_1}{K}}\breve X_2>c_2-\sqrt{\frac{k_1}{K}}\beta_0,Z_1>\beta_0)\\
&+P(\sqrt{\frac{K-k_1}{K}}\breve X_2<c_2-\sqrt{\frac{k_1}{K}}\beta_0,\sqrt{\frac{k_1}{K}}Z_1+\sqrt{\frac{K-k_1}{K}}\breve X_2>c_2)+P(Z_1>\bar\Phi^{-1}(\alpha_m))\big]\\
\leq & (1+\delta_{6m})^2\big[P(\sqrt{\frac{k_1}{K}}Z_1+\sqrt{\frac{K-k_1}{K}} Z_2>c_2,Z_1>\beta_0)\big]+(1+\delta_{6m})\sum_{j=1}^2P(Z_j>\bar\Phi^{-1}(\alpha_m))\\
\leq & (1+\delta_{6m})^2\big[P(\sqrt{\frac{k_1}{K}}Z_1+\sqrt{\frac{K-k_1}{K}} Z_2>c_2,Z_1>\beta_0)\big]+2(1+\delta_{6m})\alpha_m\\
\leq & o(\P(\sum_{i=1}^KZ'_i/\sqrt{K}>c_2))
\end{align*}

\end{proof}

\begin{proof}[Proof of Lemma \ref{lem::tail}]
\textbf{(i) Prove that (1) can leads to (2):}\par
On $\cap_{t=1}^\ell \mX^{(t)}$, 
\begin{equation*}
\sumnull \csil I(T_S<\hat t^{(\ell)})\leq 
\sumnull\csil \hat t^{(\ell)}+
\bigg\{\sum_{S\in\mB_0^{(\ell)}}\csil \hat t^{(\ell)}\bigg\}\epsilon
\end{equation*}
Combined with
\begin{equation*}
\sum_{S\in \mB_0^{(\ell)}} \csil \hat t^{(\ell)}\leq \alpha\sum_{S\in \mB^{(\ell)}} \csil\bI\{T_S<\hat t^{(\ell)}\}
\end{equation*}
and
\begin{align*}
\sum_{S\in \mB^{(\ell)}} \csil \bI\{T_S<\hat t^{(\ell)}\}&= \sumnull \csil \bI\{T_S<\hat t^{(\ell)}\}+\sumalt \csil \bI\{T_S^{(\ell)}\leq \hat t^{(\ell)}\}\\
&\leq \sumnull \csil \bI\{T_S^{(\ell)}<\hat t^{(\ell)}\}+Cm^{r_1}
\end{align*}
We have:
\begin{equation*}
(1-\alpha-\alpha\epsilon)\sumnull\csil \hat t^{(\ell)}\leq \alpha Cm^{r_1}
\end{equation*}
Thus, $2|\mB_0^{(\ell)}|\hat t^{(\ell)} \leq \sumnull\csil \hat t^{(
\ell)}\leq \frac{\alpha}{1-\alpha-
\alpha\epsilon}m^{r_1}$, for any $1\leq \ell\leq L$.\par
When $\ell=1$, by $|\mB_0^{(1)}|=m_0=m(1+o(1))$, we have $\hat t^{(\ell)}\leq Cm^{(r_1-1)}$.\par

\par
When $\ell\geq 2$, on $\cap_{k=1}^\el \mX^{(k)}$, we have 
\[\max_{k=1,...,\ell}\{FDP^{(k)}-\alpha\}<\epsilon\]
which leads to
$|\mB_0^{(\ell)}|/|\mB^{(\ell)}|\to 1$. And accordingly,
 $\hat t^{(\ell)}\leq Cm^{(r_1-1)}$.\par

\textbf{(ii) Prove that statement (2) leads to statement (3)}\par
On layer 1, $\bar \Phi(\hat c_S)=\hat t^{(1)}\leq C(m)^{r_1-1}$.
On layer $\ell\geq 2$ and $\cap_{h=1}^{\ell}\mX^{(h)}$, for all $S\in \mB^\el$,
\begin{align}
\bar \Phi(\hat c_S)
\leq G_S(\hat c_S)+\sum_{S'\in\mU(S)}\bar \Phi(\hat c_{S'})
\label{margcond3}
\end{align}
 Suppose $\bar\Phi(\hat c_{S'})\leq C(m)^{r_1-1}$ for $S'\in \cup_{k=1}^{\ell-1} \mB^{(k)}$, then together with $G_S(\hat c_S)=\hat t^\el\leq Cm^{r_1-1}$ and \eqref{margcond3}, we have
\begin{align*}
\hat c_S\geq \sqrt{2(1-r_1)\log m-2\log\log m}=\beta_m
\end{align*}
for all $S\in \mB^{(\ell)}$. \par

In addition, for $S\in\mB^\el$, on $\cap_{h=1}^{\ell-1}\mX^{(h)}$,
\begin{align}
\label{margcond1}
G_S(\hat c_S)[1-\bar\Phi(\frac{
\beta_0}{
\sqrt{M^{L-1}}})]^{M^{L-1}}\leq \bar\Phi(\hat c_S)
\end{align}
So we have $\bar \Phi(\hat c_S)\geq \hat t^{(\ell)}(1+o(1))$, and accordingly, $\hat c_S\leq \gamma_m$.\par
Note that the $\hat c_S\leq \gamma_m$ only depends on the statement (2) on layer $\ell-1$. Thus, we can apply the conclusion to show $\P(m_0\hat t^\el>c\log m)\to 1$ in the proof of theorem 1.
\par
\textbf{(iii) Prove that statement (1) holds on layer 1 ($\ell=1$):}\par
Define $\nu_m=[(|\mA'|^2/m+\delta_{2m})\vee 1]/\sqrt{c_\md\log m}$. Let $0=c_0<...<c_{\J}=\gamma_m$ satisfy $c_k-c_{k-1}=\nu_m$ for $1\leq k<\J$ and $c_{\lceil\gamma_m /\nu_m \rceil}-c_{\J-1}\leq \nu_m$. We can get the corresponding p-values sequence $q_{0}>...>q_{\J}$ with $q_k=1-\Phi(c_k)$. Let value $q^{(1)}=C^{(1)}c_\md/m$, by \eqref{mp1}, we have $P(\hat t>q^{(1)})\to 1$. We define the working p-value sequence on layer 1 as $P_{sub}^{(1)}=\{q_0,...,q_{k^{(1)}},q^{(1)}\}$, where $k^{(1)}\in\{0,...,\J-1\}$ is the index s.t. $q_{k^{(1)}}\geq q^{(1)}$ and $q_{k^{(1)}+1}\leq q^{(1)}$.

If $\forall \epsilon>0$,
\begin{align}
&P\bigg(\max_{q\in P_{sub}^{(1)}}\bigg|\frac{\sum_{S\in\mB_{0}^{(1)}}I(X_S > \bar\Phi^{-1}(q))-\sum_{S\in\mB_{0}^{(1)}}P(X_S>\bar\Phi^{-1}(q))(1-\delta_{0m})
}{\sum_{S\in\mB_{0}^{(1)}} q}\bigg|>\epsilon
\bigg)
\to 0\label{L1.1.2}
\end{align}
Then, 
\begin{align}
&P\bigg(\max_{q\in P_{sub}^{(1)}}\frac{\sum_{S\in\mB_0^{(1)}}I(T_S^{(1)}<  q)-\sum_{S\in\mB_0^{(1)}}q
}{\sum_{S\in\mB_0^{(1)}}q}>
\epsilon
\bigg)\nonumber\\
\leq & P\bigg(\max_{q\in P_{sub}^{(1)}}\frac{\sum_{S\in\mB_0^{(1)}}I(X_S>\bar\Phi^{-1}(q))-\sum_{S\in\mB_0^{(1)}}P(\tilde X_S>\bar\Phi^{-1}(q) )
}{\sum_{S\in\mB_0^{(1)}}  q}>\epsilon\bigg)\nonumber\\
\leq & P\bigg(\max_{q\in P_{sub}^{(1)}}\frac{\sum_{S\in\mB_{0}^{(1)}}I(X_S>\bar\Phi^{-1}(q) )-\sum_{S\in\mB_{0}^{(1)}}P( X_S>\bar\Phi^{-1}(q))(1-\delta_{0m})}{\sum_{S\in\mB_{0}^{(1)}}  q}>\epsilon\bigg)\nonumber\\
\leq & P\bigg(\max_{q\in P_{sub}^{(1)}}\bigg|\frac{\sum_{S\in\mB_{0}^{(1)}}I(X_S>\bar\Phi^{-1}(q) )-\sum_{S\in\mB_{0}^{(1)}}P( X_S>\bar\Phi^{-1}(q))(1-\delta_{0m})}{\sum_{S\in\mB_{0}^{(1)}}  q}\bigg|>\epsilon\bigg)\nonumber\\
=&o(1)\label{layer1}
\end{align}
 Together with the fact that $\sup_{j=1,...,k}\bigg|q_{(j)}/q_{(j-1)}-1\bigg|=o(1)$, we have 
\begin{align*}
P\bigg(\sup_{q\in [q^{(1)},\alpha]}\frac{\sum_{S\in\mB_0^{(1)}}I(T_S<  q)-\sum_{S\in\mB_0^{(1)}}q
}{\sum_{S\in\mB_0^{(1)}}q}>\epsilon\bigg)=o(1)
\end{align*}
Thus, to prove (1) holds on layer 1, we only need to show \eqref{L1.1.2}.\par
Define $C^{(1)}_{sub}=\{c_0,...,c_{k'},c'\}$, with $c'=\bar\Phi^{-1}(q')$. In order to show \eqref{L1.1.2}, it is suffice to show
\begin{equation}
\label{eq::int}
\int_0^{c'}P\bigg\{\bigg|\frac{\sum_{S\in\mB_{0}^{(1)}}I(X_S>c)-P(X_S>c)(1-\delta_{0m})}{\sum_{S\in\mB_{0}^{(1)}}\bar\Phi(c)}\bigg|\geq \epsilon\bigg\}dc=o(\nu_m)
\end{equation}
Note that by Markov inequality,
\begin{align*}
&P\bigg\{\bigg|\frac{\sum_{S\in\mB_{0}^{(1)}}\big[I(X_S>c)-P(X_S>c)(1-\delta_{0m})\big]}{\sum_{S\in\mB_{0}^{(1)}}\bar\Phi(c)}\bigg|\geq \epsilon\bigg\}\\
\leq & P\bigg\{\bigg|\frac{\sum_{S\in\mB_{0}^{(1)}}\big[I(X_S>c)-P(X_S>c)\big]}{\sum_{S\in\mB_{0}^{(1)}}\bar\Phi(c)}\bigg|\geq \epsilon-(1+\delta_{0m})\delta_{0m}\bigg\}\\
\leq &\frac{\sum_{S,S'\in\mB_{0}^{(1)}}\big[P(X_S>c,X_{S'}>c)-P(X_S>c)P(X_{S'}>c)\big]}{\big(\sum_{S\in\mB_{0}^{(1)}}\bar\Phi(c)\big)^2[\epsilon-(1+\delta_{0m})\delta_{0m}]^2}
\end{align*}
\par
We can divide the $S,S'\in\mB_0^{(1)}$ into the following three subsets:
\begin{align}
\mB_{01}^{(1)}&=\{S,S'\in\mB_0^{(1)}:S=S'\}\nonumber\\
\mB_{02}^{(1)}&=\{S,S'\in\mB_0^\el:S\neq S',\exists A,A'\in\mA^{(L)}, s.t. S\subset A, S'\subset A',\text{ and } A'\in\Gamma_A\}\label{eq::bdef}\\
\mB_{03}^{(1)}&=\{S,S'\in\mB_0^{(1)}:S\neq S'\}\setminus \mB_{02}^{(1)}\nonumber
\end{align}
Then,
\[\frac{\sum_{(S,S')\in\mB_{01}^{(1)}}\big[P(X_S>c,X_{S'}>c)-P(X_S>c)P(X_{S'}>c)\big]}{\big(\sum_{S\in\mB_{0}^{(1)}}\bar\Phi(c)\big)^2[\epsilon-(1+\delta_{0m})\delta_{0m}]^2}\leq \frac{C}{\sum_{S\in\mB_{0}^{(1)}}\bar\Phi(c)}\]
Based on condition~\ref{cond::wd},
\[\frac{\sum_{(S,S')\in\mB_{02}^{(1)}}\big[P(X_S>c,X_{S'}>c)-P(X_S>c)P(X_{S'}>c)\big]}{\big(\sum_{S\in\mB_{0}^{(1)}}\bar\Phi(c)\big)^2[\epsilon-(1+\delta_{0m})\delta_{0m}]^2}\leq \frac{C(|\mA'|^2/m+\delta_{2m})}{\sum_{S\in\mB_{0}^{(1)}}\bar\Phi(c)}\]
In addition,
\[\frac{\sum_{(S,S')\in\mB_{03}^{(1)}}\big[P(X_S>c,X_{S'}>c)-P(X_S>c)P(X_{S'}>c)\big]}{\big(\sum_{S\in\mB_{0}^{(1)}}\bar\Phi(c)\big)^2[\epsilon-(1+\delta_{0m})\delta_{0m}]^2}= o(1)\]
Thus, after some calculation, we can prove \eqref{eq::int} and then $\P(\mX^{(1)})\to 1$.
\par
Similarly, if $|\tilde \Omega_0|=m(1+o(1))$, based on \eqref{L1.1.2}, we have
\begin{align}
&P\bigg(\max_{q\in P_{sub}^{(1)}}\bigg|\frac{\sum_{S\in\mB_0^{(1)}}I(T_S^{(1)}<  q)-\sum_{S\in\mB_0^{(1)}}q
}{\sum_{S\in\mB_0^{(1)}}q}\bigg|>
\epsilon
\bigg)=o(1)\nonumber
\end{align}
Hence, $\P(\mX^{'(1)})\to 1$.


\textbf{(iv) Prove that statement (1) holds on layer $\ell\geq 2$ when statement (1) holds on previous layers:}\par

On layer $\ell$, we can divide the $S,S'\in\mB_0^\el$ into the following three subsets:
\begin{align}
\mB_{01}^\el&=\{S,S'\in\mB_0^\el:S=S',\{T_i:i\in S\} \text{ are mutually independent}\}\nonumber\\
\mB_{02}^\el&=\{S,S'\in\mB_0^\el:\exists A,A'\in\mA^{(L)}, s.t. S\subset A, S'\subset A',\text{ and } A'\in\Gamma_A\}\nonumber\\
\mB_{03}^\el&=\{S,S'\in\mB_0^\el:S\neq S'\}\setminus \mB_{02}^\el\nonumber
\end{align}
Consider the p-values sequence $q_{0}>...>q_{\J}$ constructed in (iii). Let $q^{(\ell)}=C^{(\ell)}c_\md/m$, by \eqref{mp1}, we have $P(\hat t>q^{(\ell)})\to 1$. We define the working p-value sequence on layer 1 as $P_{sub}^{(\ell)}=\{q_0,...,q_{k^{(\ell)}},q^{(\ell)}\}$, where $k^{(
\ell)}\in\{0,...,\J-1\}$ is the index s.t. $q_{k^{(\ell)}}\geq q^{(\ell)}$ and $q_{k^{(\ell)}+1}\leq q^{(\ell)}$.
\par
In view of statement (3) and Lemma \ref{lem::mcond}, we have 
\begin{align*}
\sup_{k=0,...,\J}\bigg|\frac{G_S(c_k)}{\bar \Phi(c_k)}-1\bigg|=o(1)
\end{align*}
Together with statement (3) and Lemma \ref{lem::asym}, there exists $\delta_5(m)\to 0$ with
\begin{align*}
&\max_{S\in\mB_0^\el }\frac{\P(X_S>\bar\Phi^{-1}(q)|\mQ^{(1:\ell-1)})}{q}\\
\leq& \max_{S\in\mB_{0}^\el}\frac{P(X_S>\bar\Phi^{-1}(q))}{P(Z_S>\bar\Phi^{-1}(q))[1-\bar\Phi(\frac{
\beta_0}{
\sqrt{M^{L-1}}})]^{M^{L-1}}}\\
\leq & 1+\delta_5(m)
\end{align*}

Then
$\forall \epsilon>0$,
by following the similar arguments in (iii), we can have 
\begin{align}
&P\bigg(\max_{q\in P_{sub}^{(\ell)}}\bigg|\frac{\sum_{S\in\mB_{01}^{(\ell)}}|S|I(X_S > \bar\Phi^{-1}(q))-\sum_{S\in\mB_{01}^{(\ell)}}|S|P(X_S>\bar\Phi^{-1}(q)|\mQ^{(1:\ell-1)})(1+\delta_{0m})
}{\sum_{S\in\mB_{01}^{(\ell)}} |S|q}\bigg|>\epsilon \Bigg|\mQ^{(1:\ell-1)}
\bigg)\nonumber\\
&\to 0\label{Lell.1.2}
\end{align}
Then, 
\begin{align}
&P\bigg(\max_{q\in P_{sub}^{(\ell)}}\frac{\sum_{S\in\mB_0^{(\ell)}}|S|I(T_S<  q)-\sum_{S\in\mB_0^{(\ell)}}|S|q
}{\sum_{S\in\mB_0^{(\ell)}}|S|q}>
\epsilon\Bigg|\mQ^{(1:\ell-1)}
\bigg)\nonumber\\
\leq & P\bigg(\max_{q\in P_{sub}^{(\ell)}}\frac{\sum_{S\in\mB_{0}^{(\ell)}}|S|I(X_S>\bar\Phi^{-1}(q))-\sum_{S\in\mB_{0}^{(\ell)}}|S|P( X_S>\bar\Phi^{-1}(q)|\mQ^{(1:\ell-1)} )
}{\sum_{S\in\mB_{0}^{(\ell)}}  |S|q}>\epsilon/2\Bigg|\mQ^{(1:\ell-1)}\bigg)\nonumber\\
=&o(1)\label{layerell}
\end{align}
 Together with the fact that $\sup_{j=1,...,k}\big|q_{(j)}/q_{(j-1)}-1\big|=o(1)$, we have 
\begin{align*}
P\bigg(\sup_{q\in [q^\el,\alpha]}\frac{\sum_{S\in\mB_0^{(\ell)}}|S|I(T_S<  q)-\sum_{S\in\mB_0^{(\ell)}}|S|q
}{\sum_{S\in\mB_0^{(\ell)}}|S|q}>\epsilon\Bigg|\mQ^{(1:\ell-1)}\bigg)=o(1)
\end{align*}
And thus $P(\mX^{(\ell)})\to 1$.\par
Similarly, based on Lemma~\ref{lem::mcond} (2)
, when $|\tilde\Omega_0|=m(1+o(1))$, we have $P(\mX^{'(\ell)})\to 1$.
\end{proof}

\begin{proof}[Proof of Lemma \ref{lem::equ}]
When $\ell=1$:\par
 for $\delta=1/m^4$,
\begin{align}
\sum_{S\in\mB_0^{(1)}} \csil\hat t^{(1)}&\leq \alpha\sum_{S\in\mB^{(1)}} \csil I(T_S<\hat t^{(1)})\nonumber\\
&\leq \alpha\sum_{S\in\mB^{(1)}} \csil I(T_S<\hat t^{(1)}+\delta)\nonumber\\
&\leq \sum_{S\in\mB_0^{(1)}}\csil \hat t^{(1)}(1+o(1))\label{eq::lem4p}
\end{align}

Assume \eqref{eq::lem4} holds on layer $1,\ldots,\ell-1$. Then,
\begin{equation*}
\sumnull \csil\hat t^{(\ell)}\leq \alpha(1+o(1))\sumall \csil I(T_S<\hat t^{(\ell)})
\end{equation*}
Thus, by following the similar arguments on \eqref{eq::lem4p}, we can get \eqref{eq::lem4} on layer $\ell$.

\end{proof}



\end{document}